\documentstyle[preprint,aps,eqsecnum]{revtex}
\begin{document}
\draft

\tightenlines
\newcommand{\modegy}{\,({\rm mod}\,1)}
\newcommand{\J}{{\cal J}}
\newcommand{\ch}{{\rm ch}}
\newcommand{\sh}{{\rm sh}}
\newcommand{\th}{{\rm th}}
\newcommand{\bml}{\begin{mathletters}}
\newcommand{\eml}{\end{mathletters}}
\newcommand{\bee}{\begin{equation}}
\newcommand{\ee}{\end{equation}}
\newcommand{\ba}{\begin{array}}
\newcommand{\ea}{\end{array}}
\newcommand{\bea}{\begin{eqnarray}}
\newcommand{\eea}{\end{eqnarray}}
\newcommand{\e}{\epsilon_+(i\kappa)}
\newcommand{\eps}{\epsilon}
\newcommand{\pa}{\partial}
\newcommand{\lb}{\lbrack}
\newcommand{\Se}{S_{\rm eff}}
\newcommand{\rb}{\rbrack}
\newcommand{\de}{\delta}
\newcommand{\ka}{\kappa}
\newcommand{\MSb}{{\overline {\rm MS}}}
\newcommand{\lnhb}{\ln(h/\Lambda_{{\overline {\rm MS}}})}
\newcommand{\df}{\delta f(h)}
\newcommand{\h}{{1\over2}}
\newcommand{\R}{m/\Lambda}
\newcommand\delsl{\raise.15ex\hbox{/}\kern-.57em\partial}

\def\eqalign#1{
\null \,\vcenter {\openup \jot \ialign {\strut \hfil $\displaystyle {
##}$&$\displaystyle {{}##}$\hfil \crcr #1\crcr }}\,}

\title{Scaling limit of the one-dimensional attractive  
Hubbard model: The half-filled band case}

\author{F. Woynarovich}
\address{Institute for Solid State Physics\\
of the Hungarian Academy of Sciences\\
1525 Budapest 114, Pf 49.}
\author{P. Forg\'acs}
\address{Laboratoire de Math.~et Physique Th\'eorique\\
CNRS UPRES-A 6083\\
D\'epartement de Physique\\
Facult\'e des Sciences, Universit\'e de Tours\\
Parc de Grandmont, F-37200 Tours}
\maketitle
\begin{abstract}
The scaling limit of the higher level
Bethe Ansatz (HLBA) equations for 
a macroscopically 
half-filled Hubbard chain is considered.
These equations practically decouple into
three disjoint sets which
are again of the BA type, and 
correspond to the secular equations of
three different kinds of dressed particles (one massive and two massless).
The finite size corrections and the fine structure of the 
spectrum show that the massless sector corresponds to a conformal field 
with central charge $c=1$ and Gaussian anomalous dimensions.
The zero temperature free energy is also calculated and  
is found to be in perfect agreement with the 
results of a perturbative calculation in the SU(2) chiral
Gross-Neveu (CGN) model. Some further
arguments are presented supporting the identification of 
the model obtained as the 
relativistic limit of the half-filled Hubbard chain with the SU(2) CGN model.
\end{abstract}


\section{INTRODUCTION}
\label{sec:int}
\noindent 
The one dimensional (1D) Hubbard model, being completely integrable 
\cite{Sha}
and exactly solvable by Bethe Ansatz (BA) \cite{LiWu} plays a central
role in the theory of strongly correlated electron systems \cite{Exactly}.
It is described by the Hamiltonian
\bee
\hat H = -t\sum_{j=1}^N\sum_{\sigma=\uparrow,\downarrow}
  \left( c^{+}_{j,\sigma}c^{\phantom{+}}_{j+1,\sigma}+
  c^{+}_{j+1,\sigma}c^{\phantom{+}}_{j,\sigma}\right) +
  U\sum_{j=1}^N\left( \hat n_{j,\uparrow}-{1\over2}\right)\left(
  \hat n_{j,\downarrow}-{1\over2}\right)\,,
  \label{1.1}
\ee
where $c^{+}_{j,\sigma}$ resp.\ $c^{\phantom{+}}_{j,\sigma}$ are the canonical 
electron creation resp.\ destruction operators at 
site $j$ with spin $\sigma$, and
$\hat n_{j,\sigma}=c^{+}_{j,\sigma}c^{\phantom{+}}_{j,\sigma}$.
We assume periodic boundary conditions, so the site $j=N+1$ is equivalent
to the site $j=1$. The sign of the hopping, $t$, is positive while the  
interaction, $U$, for the considered {\em attractive} case is negative.
The Hamiltonian (\ref{1.1}) has an SU(2)$\times$SU(2)
symmetry, which is in fact reduced to SO(4) due to the
selection rule that (for even $N$) only representations with {\em integer}
total spin of SU(2)$\times$SU(2) are allowed \cite{Ya,Per,EssKo}.
One of the SU(2) is associated to the spin the other we
call isospin (which is associated to the charge). 

The diagonalization of the Hamiltonian (\ref{1.1}) is reduced by the 
BA to solving 
a set of nonlinear equations \cite{LiWu}.
For a macroscopically large number of electrons the corresponding system
consists of a
macroscopically large number of equations, and to solve them 
means to reduce the original set to a smaller set  
which describes the {\em excitations\/} only. This reduced set is often 
referred to as
the higher level Bethe Ansatz (HLBA) equations, and can be considered as the
secular equations for the excitations. In this procedure also
the dispersion of the excitations is found.
The different solutions of the BA equations corresponding
to different kinds of eigenstates of the Hamiltonian (\ref{1.1}) are 
discussed in 
detail in the papers reprinted in Ref.~\cite{Exactly}.  

A {\em macroscopically
half-filled} (HF) Hubbard chain (i.e.\ in which the number of particles 
does not significantly differ from the number of sites) has two types 
of excitations \cite{Wo1,Wo2} which are connected with the two SU(2) 
symmetries of the system \cite{EssKo}. One kind of the excitations has a 
gap in the energy spectrum, the other 
is gapless. The energy and the momentum of the system is given as the 
sum of the energies and the momenta of the individual excitations.

In the present work we study the excitations (i.e.~the physical or 
dressed particles) 
in the scaling 
limit of the HF Hubbard chain. 
To gain insight into the structure of the excited states, i.e.~into
the nature of the physical particles  
of the limiting theory, 
we derive their secular equations.
The starting point of our analysis are the HLBA equations 
derived in Refs.~\cite{Wo1,Wo2}, and 
we follow the behaviour of these equations and the spectrum in the 
scaling limit. 
The scaling limit corresponds to a continuum limit together with
fine tuning the interaction strength $U$ to zero while keeping
the mass of the physical particles described by the limiting HLBA eqs.\ fixed. 
In the continuum limit the lattice spacing, $a$, tends to zero 
keeping the length of the chain $L=Na$ to be constant (so $N\to\infty$),
while the particle number per site is fixed.
In order to keep the mass scale (the energy gap)   
finite as $a\to0$ one has to adjust the hopping $t$ 
together with the dimensionless coupling $u=|U|/4t$ according to 
\bee\label{1.3}
   m_0={8t\over\pi}\sqrt{u}\,{\rm exp}\left\{-{\pi\over2u}\right\},\quad   
   t(=1/2a)\to\infty,\quad u\to0\,,
\ee
in agreement with the result found in Refs.~\cite{Ov,Me}.
As the spectrum of the Hamiltonian (\ref{1.1}) in this limit
contains both massive and massless excitations with 
{\em relativistic} dispersion relations,
it is very tempting to identify the
continuum model defined by the scaling limit
with a (1+1)-dimensional relativistic field theory.

It has been argued already a long time ago \cite{Fi} that the
scaling limit of the HF Hubbard chain can be identified with
the SU(2) symmetric chiral Gross-Neveu
\cite{GrNe} (CGN) model. The Lagrangian of the CGN model can be written as:
\bee\label{lagr1}
{\cal L} =  i\bar{\psi}\delsl\psi
- {1 \over 4} g(\bar{\psi} \gamma_{\mu} \sigma^i \psi)^2
- {1 \over 2} g'(\bar{\psi} \gamma_{\mu} \psi)^2 \,,
\ee
where $\psi$ is a doublet of Dirac fermions, and $\sigma^i$ are the Pauli
matrices.
This theory is asymptotically free \cite{MiWe} and
it belongs to the very special class
of field theories which admit 
an infinite number of conservation laws preventing 
particle production in scattering processes.
Using the `bootstrap' approach the 
$S$--matrix of the {\em massive} particles had been proposed \cite{BKKW,ABW,KoKuSw}. 
Furthermore the CGN  theory is expected to essentially decouple into a massive
and massless sector \cite{Wit}.
The Hamiltonian of (\ref{lagr1})
has been diagonalized by the BA \cite{AnLo1,Bel}.
Thereby both the proposed 
mass spectrum and the bootstrap $S$--matrix has been confirmed 
\cite{AnLo1,AnLo2}.

In Ref.~\cite{Me} both the spectrum and the
$S$ matrix of the massive particles computed in the
scaling limit of the HF Hubbard chain 
has been found to agree with those of the SU(2) CGN
model. 
The $S$-matrix of the {\em massless} sector was found to correspond to
the massless scattering theory proposed in Ref.~\cite{Zam2}
associated to a level one SU(2)$\times$SU(2) Wess-Zumino-Witten (WZW) theory.

Our approach to analyse  
the scaling limit of the HLBA equations of the HF Hubbard chain
is complementary to the existing ones \cite{Me,Fi,Aff}, and at the same time
it also makes
possible to reproduce most of the known results.
For convenience we list here our main results: 

\begin{enumerate}

\item The resulting HLBA
equations form three disjoint sets (for the massive excitations and the
right moving (`r'), resp.\ the left moving
(`l') sector)
each determining the parameters of only one kind of excitations.
(This shows that only the same type of excitations interact
in a nontrivial way.)
All the limiting HLBA eqs.\ have the same structure:
\bea
 Lp(x)&=&2\pi I_x-\sum_{x'}\phi\left({x-x'\over\pi}\right)+
  \sum_{X}2{\rm tan}^{-1}\left({x-X\over\pi/2}\right)\ ,\nonumber\\
  \sum_x 2{\rm tan}^{-1}\left({X-x\over\pi/2}\right)&=&2\pi J_X+
  \sum_{X'}2{\rm tan}^{-1}\left({X-X'\over\pi}\right)\ ,
  \label{1.5}\\
\phi(x)&=&{1\over i}{\rm ln}{\Gamma\left({1\over2}-i{x\over2}\right)
                             \Gamma\left(1+i{x\over2}\right)\over
                             \Gamma\left({1\over2}+i{x\over2}\right)
                             \Gamma\left(1-i{x\over2}\right)}\ ,\nonumber
\eea
with $p$ being the momentum of the excitations,  
$x$ stands for the rapidity
of the corresponding kind of excitations 
and the $X$s are parameters
needed to distinguish the 
various internal symmetry states of the given kind of excitations. 
The actual quantum numbers $I_x$ obey some ``parity-prescriptions". For the
massless `r'  and `l' particles these prescriptions 
depend on the number of 
particles in the given branch only, implying that the massless sector
decouples from the massive one in a sense. To be more precise the rapidities
of the `r' (`l') massless particles are independent of the
massive sector and the state of the `l' (`r') branch. (A
prescription, however, which requires that the total number of physical
particles must be even, still represents a nontrivial coupling between
the sectors.)

\item While the massive particles carry spin, the massless ones carry 
isospin (charge). We find, that for most states neither the 
`r' nor the `l'
excitations form separately isospin eigenstates,
only the `r' and `l' sectors together.
These states 
are characterized in addition to the parameters of the `r' and
`l' sectors by an additional set of parameters,
$\theta$, satisfying yet again BA-type equations:
\bee
  n_r2\tan^{-1}{\theta-2\over u}+n_l2\tan^{-1}{\theta+2\over u}
  =2\pi J_{\theta}+\sum_{\theta'}2\tan^{-1}{\theta-\theta'\over2u}\ .
  \label{1.6}
\ee
The $\theta$ parameters do not appear in the 
equations determining the rapidities  
so they do not
influence the energy, they are needed, however, to account for the isospin of
the states.
We argue that for these degenerate states, i.e.\ for those which
differ in the number and values of the $\theta$ variables only,
a new basis can be constructed 
in which both the `r' and  `l' sectors are in isospin eigenstates.
While for the original eigenstates the chiral charge is not well
defined, the new basis consists of {\em eigenstates of the
chiral charge}, i.e.~the spectrum of the limiting theory is {\em
chiral invariant}.

\item By computing finite size corrections 
 together with the fine structure of the spectrum in the scaling
limit, the conformal properties of the limiting theory are exhibited.
The  
massless sector is shown to
correspond to a conformal field with central charge $c=1$ and 
Gaussian conformal weights
\bee
\Delta={1\over4}(n+m)^2\quad{\rm and}\quad
   \bar\Delta={1\over4}(n-m)^2\,,\label{1.7}
\ee
where $n$ and $m$ are integers related to the numbers of the
`r' and `l' particles. Eq.~(\ref{1.7}) lends further support
to the identification of the massless sector with a level one WZW theory.
We note that (\ref{1.7}) together with the $S$-matrices for the
massless sectors calculated in Section III.\ also yields
convincing  evidence that the massless bootstrap $S$-matrix proposed
for level one WZW models in Ref.~\cite{Zam2} is correct.
We also note, that the theory obtained in the scaling limit is an example of a 
conformal field coupled to a massive one. We find that the spectrum of the 
massless sector has a conformal tower structure even 
if the massive sector is not
empty, but in this case the positions of the towers can be shifted.  
\item The behaviour of states with a finite density of excitations is 
also dicussed. The zero temperature free energies
(which are nothing but the ground state
energies in the presence of a chemical potential)
of the different sectors  are independent 
of each other, and the total free energy $f(\mu,\nu)$ can be written
as 
\bee 
f(\mu,\nu)=-{\mu^2\over\pi} \Psi({\mu\over m_0})-{\nu^2\over\pi}\,,
\ee
where $\mu$ and $\nu$ denote the chemical potentials for the 
massive resp.\ massless particles. We also exhibit the asymptotic series
of $\Psi({\mu/ m_0})$ (in $1/[\ln(\mu/m_0)]$) for high particle densities.
\item
Some additional evidence is provided that the scaling limit of the HF
Hubbard chain is the SU(2) CGN model (\ref{lagr1}) with $g'=0$. 
First we show that when $g'=0$  the full symmetry of the CGN model is 
in fact SO(4) which is to be expected if it really corresponds to
the scaling limit of the HF Hubbard chain.  
Secondly we show that the
zero temperature free energy calculated 
in the high density limit by perturbation theory \cite{FoNi}
is in complete agreement with
the result obtained from the scaling limit of the HLBA equations.
This also provides a rather strong (though indirect) evidence that
the beta-function
of the CGN model is correctly obtained (up to two loops) in the
scaling limit. 
We have also made a one loop renormalization group analysis of the 
{\em naive}
continuum limit of the HF Hubbard model which is neither a chiral nor
a Lorentz invariant theory. Our analysis indicates that the
renorm trajectory of the naive limit aproaches that of the CGN
model, lending further support to their identification. The one
loop result is, however, not conclusive as the renorm
trajectories are driven into the strong coupling regime.

\end{enumerate}  

The paper is organized as follows. First (Sec.\ \ref{sec:hlba}.) we review 
shortly the BA 
and HLBA equations describing the states in 
question. The relativistic limit of these equations together with their 
interpretation is given in the 
second part (Sec.\ \ref{sec:rellim}.), while
the third part (Sec.\ \ref{sec:conform}.) is devoted to the 
discussion of the finite size 
corrections and conformal properties.
In the fourth part (Sec.\ \ref{sec:dens}.) some states with a finite 
density of excitations are discussed. In the last part 
(Sec.\ \ref{sec:cgn}.) some arguments supporting the equivalence
of the SU(2) (actually SO(4)) 
symmetric CGN model and the scaling limit of the
half filled Hubbard 
chain are collected. 
At the end of the paper in Appendix \ref{sec:A} 
we illustrate the
isospin structure of the massless states. 
In Appendix \ref{sec:delta} we show that the approximations used to
derive the HLBA equations for the discrete chain remain correct in the 
scaling limit too,
and in Appendix \ref{sec:vegesm} the finite size corrections are discussed.
In Appendix \ref{sec:fren} we sketch
the perturbative calculation of the free energy in the CGN
model, while Appendix \ref{sec:nlim} contains the derivation of the naive
continuum limit of HF Hubbard chain. Finally in Appendix \ref{sec:coupl} 
a one--loop
renormalization group analysis of the theory obtained in the 
naive continuum limit is given.

\section{THE BETHE ANSATZ EQUATIONS}
\label{sec:hlba}
\noindent
Let us exhibit first the two commuting SU(2) 
symmetries of the Hubbard Hamiltonian (\ref{1.1})
explicitly \cite{Ya,EssKoSchou}: 
the `spin' SU(2) is generated by 
\begin{mathletters}
\bea
S^x&=&{1\over2}\sum_j(c^{+}_{j,\uparrow}
c^{\phantom{+}}_{j,\downarrow}
+c^{+}_{j,\downarrow}c^{\phantom{+}}_{j,\uparrow})\,,\nonumber\\ 
S^y&=&{1\over2i}\sum_j(c^{+}_{j,\uparrow}c^{\phantom{+}}_{j,\downarrow}
-c^{+}_{j,\downarrow}c^{\phantom{+}}_{j,\uparrow})\,,\label{1.2a}\\
S^z&=&{1\over2}\sum_j(\hat n_{j,\uparrow}-\hat n_{j,\downarrow})\,,\nonumber
\eea
while the generators of the other SU(2) (`isospin') which is related to 
the charge are:
\bea
I_1&=&{1\over2}\sum_j(-1)^j(c^{\phantom{+}}_{j,\uparrow}
c^{\phantom{+}}_{j,\downarrow}
+c^{+}_{j,\downarrow}c^{+}_{j,\uparrow})\,,\nonumber\\
I_2&=&{1\over2i}\sum_j(-1)^j(c^{\phantom{+}}_{j,\uparrow}
c^{\phantom{+}}_{j,\downarrow}
-c^{+}_{j,\downarrow}c^{+}_{j,\uparrow})\,,\label{1.2b}\\ 
I_3&=&{1\over2}\sum_j(1-\hat n_{j,\uparrow}-\hat n_{j,\downarrow})\,.\nonumber
\eea
We remark that the spin and the isospin can be mapped onto each other by 
the transformation 
\bee
\begin{array}{rl}
c^{\phantom{+}}_{j,\downarrow}&\to 
c^{\phantom{+}}_{j,\downarrow}\,,\\
c^+_{j,\downarrow}&\to c^+_{j,\downarrow}\,,\end{array}\qquad
\begin{array}{rl}c^{\phantom{+}}_{j,\uparrow}&\to(-1)^j c^+_{j,\uparrow}\,,\\
c^+_{j,\uparrow}&\to(-1)^j c^{\phantom{+}}_{j,\uparrow}\,.
\end{array}\label{1.2c}
\ee
\end{mathletters} 
(This transformation applied to the $\hat H$ changes the sign of $U$.)

The diagonalizing the (\ref{1.1}) Hamiltonian is equivalent to solving the set
of nonlinear equations \cite{LiWu}
\begin{mathletters}\label{2.1}
\bee
Nk_j=2\pi I_j-\sum_{\alpha = 1}^M2\tan^{-1}{\sin k_j-\lambda_{\alpha}
 \over U/4t}\ ,\label{2.1a}
\ee
\bee
\sum_{j=1}^{N_e}2\tan^{-1}{\lambda_{\alpha}-\sin k_j\over U/4t}=
 2\pi J_{\alpha}+ \sum_{\beta=1}^M 2\tan^{-1}{\lambda_{\alpha}-
 \lambda_{\beta}\over U/2t}\ .\label{2.1b}
\ee
\end{mathletters}
Here $N_e(\leq N)$ is the number of electrons and $M(\leq N_e/2)$ is the 
number of down spins, 
i.e. $S^z=(N_e/2-M)$ and $I_3=(N-N_e)/2$. The solutions correspond to 
longest spin- and isospin-projection states ($S^2=S^z(S^z+1)$, 
$I^2=I_3(I_3+1)$)  
provided all the $\lambda$ are (whatever large but) finite. 
The $I_j$ and $J_{\alpha}$ quantum numbers are integers or 
half-odd-integers depending on the parities of $N_e$ and $M$:
$I_j=M/2\modegy$, $J_{\alpha}=(N_e+M+1)/2\modegy$.
Once equations (\ref{2.1}) are solved the energy and the momentum of the 
corresponding state can be calculated:
\bee
 E={NU/4}-\sum_{j=1}^{N_e}(2t\cos k_j+U/2)\,,\quad\quad
 P=\sum_{j=1}^{N_e}k_j\ ,\label{2.2}
\ee
and also the wave-function can be given \cite{Wo3}. 

For the  considered $U<0$ attractive chain, near the ground-state 
most of the electrons
form bound pairs with wavenumbers given 
(up to corrections exponentially small in the chain length (see Appendix
\ref{sec:delta})) as  
\bee
 \sin k^{\pm}=\Lambda\pm iu\,,\label{2.3}
\ee
with $u=|U|/4t$ as given earlier, and ${\Lambda}$ being a subset of the set 
${\lambda}$. 
By this relation $k^{\pm}$ can be eliminated from Eqs.\ (\ref{2.1a}) and
(\ref{2.1b})
and one finds that the $k$ wavenumbers of the unbound electrons, the $\lambda$s
connected with the distribution of their (uncompensated) spins 
(i.e.\ those elements of the original $\lambda$ set which are not in the 
subset $\Lambda$),
and the $\Lambda$s of the bound pairs
satisfy the equations \cite{Wo2}
\bml\label{2.4}
\bea
2\pi I_j&=&Nk_j - \nonumber\\
  &&-\sum_{\alpha = 1}^{n(\lambda)}
 2\tan^{-1}{\sin k_j-\lambda_{\alpha}\over u}-
 \sum_{\eta = 1}^{n(\Lambda)}
 2\tan^{-1}{\sin k_j-\Lambda_{\eta}\over u}\ ,\label{2.4a}\\
 I_j&=&{n(\lambda)+n(\Lambda)\over2}\,\modegy\ ;\nonumber
\eea
\bee
\ba{c}{\displaystyle
\sum_{j=1}^{n(k)}2\tan^{-1}{\lambda_{\alpha}-\sin k_j\over u}=
 2\pi J_{\alpha}+ \sum_{\beta=1}^{n(\lambda)}2\tan^{-1}{\lambda_{\alpha}-
 \lambda_{\beta}\over2u}}\ ,\\
 {\displaystyle J_{\alpha}={n(k)-n(\lambda)+1\over2}\,\modegy}\ ;\quad
\ea
 \label{2.4b}
\ee
\eml
\bea
2\pi J_{\eta}&=&N\left(\sin^{-1}\!(\Lambda_{\eta}-iu)+
  \sin^{-1}\!(\Lambda_{\eta}+iu)\right)\,-\nonumber\\
  &&-\sum_{j=1}^{n(k)}2\tan^{-1}{\Lambda_{\eta}-\sin k_j\over u}
  -\sum_{\nu=1}^{n(\Lambda)}2\tan^{-1}{\Lambda_{\eta}-
  \Lambda_{\nu}\over 2u}\ ,
  \label{2.5}\\
  J_{\eta}&=&{n(k)+n(\Lambda)+1\over2}\,\modegy\ .\nonumber
\eea  
Here $n(k)$, $n(\lambda)$ and $n(\Lambda)$ denote the number of unbound electrons,
the number of unbound electrons with down spins and the number of bound pairs,
respectively ($N_e=n(k)+2n(\Lambda)$, $M=n(\lambda)+n(\Lambda)$). (We display
the prescriptions for the quantum numbers $I$ and $J$ together with the 
equations: 
in the following derivations the quantum numbers will be redefined by 
absorbing different constants into them and so  
the actual prescriptions will change from step to step.) 
The energy and momentum expressed in terms of these variables are
\bee
\ba{l}
  {\displaystyle E=-Nu-\sum_j 2t(\cos k_j-u)-}\\
  {\displaystyle \phantom{E=}-\sum_{\eta}2t\left(\sqrt{1-(\Lambda_{\eta}-iu)^2}
  +\sqrt{1-(\Lambda_{\eta}+iu)^2}-2u\right)}\ ,\\
  {\displaystyle P={2\pi\over N}
  \biggl(\sum_jI_j-\sum_{\alpha}J_{\alpha}+\sum_{\eta}J_{\eta}
  \biggr)^{\vphantom{l}}}\ .
\ea
\label{2.6}
\ee

These equations can be solved using methods by now 
standard in the treatment
of BA type equations \cite{Wo1,KluSch,DeLo1}. The applicaton
of these methods for the present case involve the 
following considerations and steps: 

\begin{itemize}
\item[i.] In the ground state all of the $N_e=N$ 
(half filling) electrons are condensed into bound pairs, whose rapidities  
are given by Eq.\ (\ref{2.5}) with $n(k)=0$ and $J_{\eta}$ being consecutive
integers or half odd integers between $\pm(N/2-1)/2$. For a macroscopically 
large $N$ the $\Lambda_{\eta}$s satisfying Eqs.\ (\ref{2.5}) will
be given by a distribution, $\sigma_0(\Lambda)$, 
so that the number of the $\Lambda_{\eta}$s within the interval 
$(\Lambda,\Lambda+d\Lambda)$ is given by $N\sigma_0(\Lambda)d\Lambda$. 
Then the summations over 
the $\Lambda$s in (\ref{2.5}) and in (\ref{2.6}) for the energy can be
replaced by integrals: 
$\sum_{\nu}f(\Lambda_{\nu})\to\int f(\Lambda)\sigma_0(\Lambda)d\Lambda$
and the density, $\sigma_0(\Lambda)$, is determined by a linear integral equation 
derived from (\ref{2.5}). 

\item[ii.] There can be two typs of excitations. One is connected with 
breaking up pairs, that is introducing
real $k$s and $\lambda$s. The other type is introduced by changing the 
distribution of $\Lambda$s by leaving holes in the ground state distribution 
and introducing complex $\Lambda$s. Also for the excited states
the set of real $\Lambda$s
can be described by a density, $\sigma(\Lambda)$, differing from the 
ground state density in $O(1/N)$ terms due to the excitations. After 
eliminating the real $\Lambda$s from the set (\ref{2.4}a-b) and 
(\ref{2.5}) one is left
with a set of equations for the unknown
$k$s (which are real) and $\lambda$s, 
the positions of the holes in the real $\Lambda$
distribution, $\Lambda_h$, and
the set of variables $\Theta_n$ representing the complex $\Lambda$s
(Appendix \ref{sec:delta}). The 
HLBA equations obtained this way read:
\end{itemize}
\bml\label{2.7}
\bee\ba{l}
  {\displaystyle N\Bigg(k_j-\int\limits^{\infty}_{-\infty}{e^{-2|\omega|u}\over
  1+e^{-2|\omega|u}}J_0(\omega)\sin(\omega\sin k_j){ d\omega\over\omega}
  \Bigg)=2\pi I_j+}\\
  \quad{\displaystyle \sum_{\alpha}2\tan^{-1}{\sin k_j-\lambda_{\alpha}\over u}-
  \sum_{j'}\phi\biggl({\sin k_j-\sin k_{j'}\over2u}\biggr)-}\\
  \quad{\displaystyle \sum_h2\tan^{-1}\biggl(\th{\pi(\sin k_j-\Lambda_h)\over4u}
  \biggr)\ ,}\\
  \ \ {\displaystyle I_j={N-n(h)-(n(k)-2n(\lambda))\over4}\modegy}\ ;
\ea
\label{2.7a}
\ee
\bee\ba{l} 
{\displaystyle \sum_{j}2\tan^{-1}{\lambda_{\alpha}-\sin k_j\over u}=
 2\pi J_{\alpha}+ \sum_{\beta}2\tan^{-1}{\lambda_{\alpha}-
 \lambda_{\beta}\over2u}\ ,}\\
 \ {\displaystyle J_{\alpha}={n(k)+n(\lambda)+1\over2}\modegy}\ ;
\ea
\label{2.7b}
\ee
\eml\bml\label{2.8}
\bee
\ba{l}
 {\displaystyle N\int\limits^{\infty}_{-\infty}{J_0(\omega)\over2\ch\omega u}
 \sin\omega\Lambda_h{ d\omega\over\omega}=2\pi J_h+
 \sum_n2\tan^{-1}{\Lambda_h-\Theta_n\over u}-}\\
 \quad{\displaystyle \sum_{h'}\phi\left({\Lambda_h-\Lambda_{h'}\over2u}\right)+
 \sum_j 2\tan^{-1}\biggl(\th{\pi(\Lambda_h-\sin k_j)\over4u}\biggr)}\ ,\\
 \ {\displaystyle J_h={1\over2}\left({N+n(k)+2n(\Theta)-n(h)\over2}+1\right)
 \modegy}\ ;
 \ea
 \label{2.8a}
\ee
\bee\ba{l}
 {\displaystyle \sum_h2\tan^{-1}{\Theta_n-\Lambda_h\over u}=2\pi J_n+
 \sum_{n'}2\tan^{-1}{\Theta_n-\Theta_{n'}\over2u},}\\
 \ {\displaystyle J_n={n(h)+n(\Theta)+1\over2}\modegy}\ .
\ea\label{2.8b}
\ee
\eml
Here $n(h)$ denotes the number of holes (i.e.\ the number of $\Lambda_h$) and
$n(\Theta)$ is the number of variables $\Theta$. 
The energy and the momentum of the state given by the solution of 
Eqs.\ (\ref{2.7}a-b)
and (\ref{2.8}a-b) are
\bml\label{2.9}
\bee
E=N\varepsilon_0+\sum_j \varepsilon_s(k_j)+\sum_h\varepsilon_c(\Lambda_h)
\ ,\label{2.9a}
\ee
\bee
P=\sum_j p_s(k_j)+\sum_h p_c(\Lambda_h)\ ,\label{2.9b}
\ee
\eml
with
\bml\label{2.10}
\bea
\varepsilon_0&=&-2t\int\limits_{-\infty}^{\infty}
 {J_1(\omega)J_0(\omega)e^{-|\omega|u}\over2\ch\omega u}{ d\omega\over\omega}
  -tu\ ,\nonumber\\
  \varepsilon_s(k)&=&-2t\Bigg((\cos k-u)-\int\limits_{-\infty}^{\infty}
  {J_1(\omega)e^{-|\omega|u}\over2\ch\omega u}\cos(\omega\sin k)
  { d\omega\over\omega}\Bigg)\ ,\label{2.10a}\\ 
  \varepsilon_c(\Lambda)&=&2t\int\limits_{-\infty}^{\infty}
 {J_1(\omega)\over2\ch\omega u}\cos\omega\Lambda{ d\omega\over\omega}\ ,
\nonumber
\eea
and
\bee\ba{rl}
  {\displaystyle p_s(k)}&
  {\displaystyle =\Bigg(k-\int\limits_{-\infty}^{\infty}
  {J_0(\omega)e^{-|\omega|u}\over2\ch\omega u}\sin(\omega\sin k)
  { d\omega\over\omega}\Bigg)}\ ,\\ 
  {\displaystyle p_c(\Lambda)}&
  {\displaystyle =-\int\limits_{-\infty}^{\infty}
 {J_0(\omega)\over2\ch\omega u}\sin\omega\Lambda{ d\omega\over\omega}}\ .
\ea
 \label{2.10b}
\ee
\eml
(This result is the same as the one obtained from the repulsive
case \cite{Wo1} using the
complementary solutions \cite{Wo2} of the Eqs.\ (\ref{2.1}a-b). 
The convention used for the
indices $s$ and $c$ is that of Ref.\ \cite{Me}.) 
The prescriptions for the $I$ and $J$ quantum numbers are expressed by the 
numbers of excitations only and numbers like $n(\Lambda)$ are
eliminated. This is possible through the relation 
\bee
 n(h)=N-2n(\Lambda)-n(k)+2n(\Theta)\,,\label{2.11}
\ee
which is a consequence of Eq.\ (\ref{2.5}). This relation also
implies, that (for even $N$) the 
number of excitations $n(k)+n(h)$ is always even.
It also allows to express the spin and the isospin (charge) in a simple form
\bee\ba{ll}
  S^z={\displaystyle{n(k)-2n(\lambda)\over2_{\vphantom{j}}}}\ ,
  \quad&\big(S^2=S^z(S^z+1)\big)\ ,\\
  I_3={\displaystyle{n(h)-2n(\Theta)^{\vphantom{2}}\over2}}\ ,
  \quad&\big(I^2=I_3(I_3+1)\big)\ ,
\ea
  \label{2.12}
\ee
which suggests, that the two kind of excitations carry either spin or isospin
(charge) but not both. It is to be emphasized, that while both $S^z$ and $I_3$ are either integers 
or half-odd-integers, due to the above requirement their sum is always an 
integer. (After all this is a consequence of the fact that changing the number
of electrons in the system by one will change both $S^z$ and $I_3$ by
plus or minus 1/2.) 

\section{THE RELATIVISTIC LIMIT}
\label{sec:rellim}

The HLBA equations and the energy-momentum dispersions have the form of 
(\ref{2.7}a-b), (\ref{2.8}a-b) 
and (\ref{2.10}a-b) in the half filled case only, 
i.e.~using them automatically fixes the band filling.   
Since letting $N\to\infty$ means also $N_e\to\infty$, to keep
the mass gap
finite the interaction $u$ has to $\to0$.   
In the $u\to0$ limit \cite{Me,Fi}
\bml\label{3.1}
\bee\ba{l} 
  \varepsilon_s(k)=(8t/\pi)K_1(\pi/2u)\ch(\pi\sin k/2u)\ ,\\
  p_s(k)=(4/\pi)K_0(\pi/2u)\sh(\pi\sin k/2u)\ ,
\ea\label{3.1a}
\ee
and
\bee\ba{l}
  {\displaystyle \varepsilon_c(\Lambda)={4t}I_1(\pi/2u)
  e^{-\pi|\Lambda|/2u}}\ ,\\
  {\displaystyle p_c(\Lambda)={\rm sgn}\Lambda\left(-{\pi\over2}+{2}I_0(\pi/2u)
  e^{-\pi|\Lambda|/2u}\right)}\ ,\\
  \ \ {\displaystyle (|\Lambda|>1)}\ ,\ea\label{3.1b}
\ee 
\eml
with $K_{0,1}$ and $I_{0,1}$ being the modified Bessel functions. The 
lattice- or quasi-momenta $p_s$ and $p_c$ are dimensionless as they generate 
the discrete translations of
a lattice. To obtain the corresponding momenta in the continuum, one has to
divide them by the lattice spacing $a$.
In the $a\to0$ limit, however, the $\pm\pi/2a$ terms in $p_c/a$
would diverge. For this reason, before taking the $a\to0$ limit the 
lattice has to be 
redefined in such a way, that four lattice sites make one elementary cell. 
Then 
the quasi-momenta $\pm\pi/2$ are {\em equivalent} to zero, i.e. they can be
dropped from the total momentum. Taking now the limit (\ref{1.3}) leads 
to 
\bml\label{3.2}
\bee\ba{r}
E-N\varepsilon_0=\sum\epsilon_m(\kappa)+
  \sum\epsilon_r(\eta)+\sum\epsilon_l(\xi)\ ,\\
  P=\sum p_m(\kappa)+\sum p_r(\eta)+ \sum p_l(\xi)\ ,\ea\label{3.2a}
\ee
with
\bee\ba{lll}
  {\displaystyle \epsilon_m(\kappa)=\lim\varepsilon_s(k)},&
  {\displaystyle p_m(\kappa)=\lim p_s(k)/a},&
  {\displaystyle \kappa={\pi\sin k\over2u}};\\
  {\displaystyle \epsilon_r(\eta)=
  \lim\left.\varepsilon_c(\Lambda)\right\vert_{\Lambda>1}},&
  {\displaystyle p_r(\eta)=
  \lim\left.{p_c(\Lambda)+\pi/2\over a}\right\vert_{\Lambda>1}},&
  {\displaystyle \eta=-{\pi(\Lambda-2)\over2u}};\\
  {\displaystyle \epsilon_l(\xi)=
  \lim\left.\varepsilon_c(\Lambda)\right\vert_{\Lambda<-1}},&
  {\displaystyle p_l(\xi)=
  \lim\left.{p_c(\Lambda)-\pi/2\over a}\right\vert_{\Lambda<-1}},&
  {\displaystyle \xi=-{\pi(\Lambda+2)\over2u}};\ea
  \label{3.2b}
\ee
\eml
$\epsilon_{m(r,l)}$ and $p_{m(r,l)}$ given by 
\bee
\ba{rlrl}
  {\displaystyle \epsilon_m(\kappa)}&
  {\displaystyle =m_0\cosh(\kappa)}\,,\quad&
  {\displaystyle p_m(\kappa)}&
  {\displaystyle =m_0\sinh(\kappa)}\,,\\ 
  {\displaystyle \epsilon_r(\eta)}&=
  {\displaystyle {m_0\over2}\exp(+\eta)}\,,&
  {\displaystyle p_r(\eta)}&
  {\displaystyle =+{m_0\over2}\exp(+\eta)}\,,\\
  {\displaystyle \epsilon_l(\xi)}&
  {\displaystyle ={m_0\over2}\exp(-\xi)}\,,&
  {\displaystyle p_l(\xi)}&
  {\displaystyle =-{m_0\over2}\exp(-\xi)}\,.
\ea  
  \label{1.4}
\ee
As one can see from (\ref{1.4}) and (\ref{3.2}a-b) in the $u\to0$ limit
the energies and momenta 
are finite, if  $\kappa$, $\eta$ and $\xi$ stay finite, i.e.
$k\to0$ and $\Lambda\to+$or$-2$. The limits of the HLBA equations are 
compatible with these requirements.

The HLBA equations for the massive particles are obtained from 
Eqs.\ (\ref{2.7a}) and ({\ref{2.7b})
by substituting (\ref{2.10b}), (\ref{3.2b}) and 
$\chi_{\alpha}=\pi\lambda_{\alpha}/2u$
into them:
\bee\ba{l}
  {\displaystyle Lp_m(\kappa_j)=2\pi I_j+\sum_{\alpha}2\tan^{-1}
  {\kappa_j-\chi_{\alpha}\over\pi/2}-
  \sum_{j'}\phi\left({\kappa_j-\kappa_{j'}\over\pi}\right)-}\\
  {\displaystyle \phantom{L}\sum_f2\tan^{-1}
  \left(\th\left({\kappa_j+\eta_f\over2}-{\pi\over2u}\right)
  \right)
  -\sum_q2\tan^{-1}\biggl(\th\biggl({\kappa_j+\xi_q\over2}+{\pi\over2u}\biggr)
  \biggr)}\ ,\\
  {\displaystyle \phantom{L}I_j={N-n(h)-(n(k)-2n(\lambda))\over4}\modegy}\ .\ea
  \label{3.3}
\ee
The last two terms give a constant
\bee\ba{l}
 {\displaystyle \lim\Bigg(\sum_f2\tan^{-1}\left(\th\left({\kappa_j+\eta_f\over2}-
  {\pi\over2u}\right)\right)+}\\
 {\displaystyle \phantom{\lim}
 +\sum_q2\tan^{-1}\biggl(\th\biggl({\kappa_j+\xi_q\over2}+
 {\pi\over2u}\biggr)\biggr)\Bigg)
  ={\pi\over2}\big(n(\xi)-n(\eta)\big)}\,,\ea\label{3.4}
\ee
with $n(\eta)$ resp.\  $n(\xi)$ denoting the number of $\eta$s resp.\ $\xi$s.
The constant (\ref{3.4}) can be absorbed into $I_j$ (from which  
$N/4$ can be dropped, as due to the redefinition of the lattice, 
$N$ must be a multiple of 4). Finally 
using that $n(\eta)+n(\xi)=n(h)$ and renaming $n(k)$ and $n(\lambda)$ to
$n(\kappa)$ and $n(\chi)$ respectively, one obtains
\bml\label{3.5}
\bee\ba{l}
 {\displaystyle Lp_m(\kappa_j)=2\pi I_j+\sum_{\alpha}2\tan^{-1}
  {\kappa_j-\chi_{\alpha}\over\pi/2}-
  \sum_{j'}\phi\left({\kappa_j-\kappa_{j'}\over\pi}\right)}\\
 \ \ {\displaystyle I_j=\pm{n(\eta)+n(\xi)+(n(\kappa)-2n(\chi)^{\vphantom{l}})
 \over4_{\vphantom{j}}}+
 {n(\eta)-n(\xi)\over4}\modegy}\\
 \ \ {\displaystyle \phantom{I_j}=\pm{n_r+n_l+(n(\kappa)-
 2n(\chi))^{\vphantom{l}}\over4}+
 {n_r-n_l\over4}\modegy}\ ;\ea\label{3.5a}
\ee 
(with $n_{r,l}$ defined later in (\ref{3.8a}) and (\ref{3.9a})), and
\bee\ba{l}
 {\displaystyle \sum_{j}2\tan^{-1}{\chi_{\alpha}-\kappa_j\over \pi/2}=
 2\pi J_{\alpha}+ \sum_{\beta}2\tan^{-1}{\chi_{\alpha}-
 \chi_{\beta}\over\pi}}\ ,\\
 \ {\displaystyle J_{\alpha}={n(\kappa)+n(\chi)+1\over2}\modegy}\ .
 \ea\label{3.5b}
\ee
\eml
We have written $I_j$ in this form (using that the first term is either an 
integer
or a half-odd-integer) to emphasize its 
symmetry with respect to $n(\eta)$ and $n(\xi)$ ($n_r$ and
$n_l$). This is actually
the left-right symmetry of the system: interchanging 
$n(\eta)$ and $n(\xi)$, and changing the sign of all $\kappa$ and $\chi$
yields an admissible solution (i.e.\ with proper quantum numbers).   

The limit of the HLBA equations for the massless particles is constructed 
in a similar way. First we observe that as $u\to0$
Eqs.\ (\ref{2.8a}) force the $\Lambda_h$s to
take their values around 2 and $-$2. At the same time 
some of the $\Theta$ condense around 2 and $-$2, but some may have 
values $|\Theta|\not\simeq2$. So the set of $\Theta$ can be replaced by 
three sets:
\bee\ba{l}
  \vartheta= -{\pi(\Theta-2)/2u}\quad(\Theta\sim2),\\
  \varphi= -{\pi(\Theta+2)/2u}\quad(\Theta\sim-2),\\
  \theta=\Theta\quad\bigl(\lim(2-|\Theta|)\neq0\bigr)\ .\ea
  \label{3.6}
\ee
(Their numbers satisfy 
$n(\vartheta)+n(\varphi)+n(\theta)=n(\Theta)$).
In the variables (\ref{3.6}), equations (\ref{2.8a}) and (\ref{2.8b})
split into three disjunct sets which can 
be treated in the same way as the set describing the massive particles. 
For the right-going particles ($\eta$ and $\vartheta$) one has 
\bml\label{3.7}
\bee\ba{l}
 {\displaystyle N\biggl({\pi\over2}-ap_r(\eta_f)\biggr)
 =2\pi J_f-
 \sum_g2\tan^{-1}{\eta_f-\vartheta_g\over\pi/2}+
 \sum_{f'}\phi\biggl({\eta_f-\eta_{f'}\over\pi}\biggr)}\\
 \ {\displaystyle -\sum_k2\tan^{-1}\biggl({\eta_f-\varphi_k\over\pi/2}-
 {4\over u}\biggr)
 -\sum_n2\tan^{-1}\biggl({\eta_f\over\pi/2}-{2-\theta_n\over u}\biggr)}\\
 \ {\displaystyle +\sum_q\phi\biggl({\eta_f-\xi_q\over\pi}-{2\over u}\biggr)
 -\sum_j 2\tan^{-1}\biggl(\th\biggl({\eta_f+\kappa_j)\over4u}-{\pi\over2u}
 \biggr)\biggr)}\ ,\\
 \ {\displaystyle J_f={1\over2}\left({N+n(\kappa)+2n(\vartheta)-n(\eta)
 +2n(\varphi)-n(\xi)+2n(\theta)\over2}+1\right)\modegy}\ ;\ea
 \label{3.7a}
\ee
\bee\ba{c}
 {\displaystyle -\sum_f 2\tan^{-1}{\vartheta_g-\eta_f\over\pi/2}-
        \sum_q2\tan^{-1}\biggl({\vartheta_g-\xi_q\over\pi}-{4\over u}\biggr)
        =2\pi J_g-}\\
 {\displaystyle \sum_{g'}2\tan^{-1}{\vartheta_g-\vartheta_{g'}\over\pi}
 -\sum_{k}2\tan^{-1}\biggl({\vartheta_g-\varphi_{k}\over\pi}-{2\over u}\biggr)}
 \\
 {\displaystyle -\sum_{n}2\tan^{-1}\biggl({\vartheta_g\over\pi}-
 {2-\theta_n\over2u}\biggr)} ,\\
 {\displaystyle J_g={n(\eta)+n(\xi)+n(\vartheta)+n(\varphi)+n(\theta)+1\over2}
 \modegy}\ .\ea
 \label{3.7b}
\ee
\eml
By absorbing all those terms into 
$J_f$ and $J_g$, which become constants in the $u\to0$ limit 
(and using that $n(\kappa)+n(\xi)+n(\eta)=even$) leads to
\bml\label{3.8}
\bee\ba{c}
 {\displaystyle Lp_r(\eta_f)=2\pi J_f+
 \sum_g2\tan^{-1}{\eta_f-\vartheta_g\over\pi/2}-
 \sum_{f'}\phi\left({\eta_f-\eta_{f'}\over\pi}\right)\,,}\\
 {\displaystyle J_f=-\left({n_r\over4}+{1\over2}\right)\modegy\ ,
 \quad n_r=n(\eta)-2n(\vartheta)}\ ;\ea
 \label{3.8a}
\ee
\bee\ba{c}
 {\displaystyle \sum_f2\tan^{-1}{\vartheta_g-\eta_f\over\pi/2}
 =2\pi J_g+\sum_{g'}2\tan^{-1}{\vartheta_g-\vartheta_{g'}\over\pi}}\ ,\\
 {\displaystyle J_g={n(\eta)+n(\vartheta)+1\over2}\modegy}\ .\ea\label{3.8b}
\ee
\eml
An analogous procedure yields for the left-going particles 
($\xi$ and $\varphi$)
\bml\label{3.9}
\bee\ba{c}
 {\displaystyle Lp_l(\xi_q)=2\pi J_q+
 \sum_k2\tan^{-1}{\xi_q-\varphi_k\over\pi/2}-
 \sum_{q'}\phi\biggl({\xi_q-\xi_{q'}\over\pi}\biggr)}\,,\\
 {\displaystyle J_q=\left({n_l\over4}+{1\over2}\right)\modegy\ ,
 \quad n_l=n(\xi)-2n(\varphi)}\ ;\ea
 \label{3.9a}
\ee
\bee\ba{c}
 {\displaystyle \sum_q2\tan^{-1}{\varphi_k-\xi_q\over\pi/2}
 =2\pi J_k+\sum_{k'}2\tan^{-1}{\varphi_k-\varphi_{k'}\over\pi}}\ ,\\
 {\displaystyle J_k={n(\xi)+n(\varphi)+1\over2}\modegy}\ .\ea\label{3.9b}
\ee
\eml
Finally the variables $\theta$ are defined by the $u\to0$ limit of the 
solution for the system
\bee\ba{l}
  {\displaystyle n(\eta)2\tan^{-1}{\theta_n-2\over u}-
          n(\vartheta)2\tan^{-1}{\theta_n-2\over 2u_{\vphantom{j}}}+}\\
  {\displaystyle n(\xi)2\tan^{-1}{\theta_n+2\over u}-
          n(\varphi)2\tan^{-1}{\theta_n+2^{\vphantom{l}}\over 2u}
  =2\pi J_n+\sum_{n'}2\tan^{-1}{\theta_n-\theta_{n'}\over2u}}\ ,\\
 \quad\quad {\displaystyle J_n=
 {n(\eta)+n(\xi)+n(\vartheta)+n(\varphi)+n(\theta)+1\over2}
 \modegy}\ .
 \ea\label{3.10}
\ee
As for $\theta\not\approx\pm2$ 
\bee
2\tan^{-1}{\theta\pm2\over 2u}=4\tan^{-1}{\theta\pm2\over u}\mp
\pi+O(u^2)\ ({\rm mod}\ 2\pi)\,,\label{3.11}
\ee
the $u\to0$ limit of solutions of (\ref{3.10}) is the same
as the $u\to0$ limit of 
solutions for the equation 
\bee\ba{c}
  {\displaystyle n_r2\tan^{-1}{\theta_n-2\over u}+n_l2\tan^{-1}{\theta_n+2\over u}
  =2\pi J_n+\sum_{n'}2\tan^{-1}{\theta_n-\theta_{n'}\over2u}}\ ,\\
  {\displaystyle J_n={n_r+n_l+n(\theta)+1\over2}\modegy}\ .
  \ea\label{3.12}
\ee
We remark that Eqs.\ (\ref{3.12}) have the same structure as
the BA equation of
the chiral invariant Gross-Neveu model \cite{AnLo1}. We also note that
the $\theta$s do not influence the energy (not even 
indirectly through some other quantity). They cannot be discarded, however, 
as they are important to give the isospin-state corresponding to a solution.

Equations (\ref{3.5}a-b), (\ref{3.8}a-b), (\ref{3.9}a-b) and 
(\ref{3.12}) are the scaling limits 
of Eqs.\ (\ref{2.7}a-b) and (\ref{2.8}a-b), and they are the secular 
equations of the
limiting model. Their solutions are meaningful under 
certain conditions. Firstly all roots have to be 
different: if two $k$s, $\lambda$s or two $\vartheta$s, $\theta$s or
$\varphi$s are equal, the wavefunction is identically zero, thus such 
solutions are (although formally possible) meaningless. 
Similarly no two
$\eta$s or $\xi$s can be equal as these variables are originally holes
in the $\Lambda$ distribution. 
Based on the fact that (\ref{3.5b}), (\ref{3.8b}),
(\ref{3.9b}) and (\ref{3.12}) are very similar in structure to the 
original equations (\ref{2.1b}),
we expect them to have meaningful solution if 
$n(\eta)-2n(\vartheta)=n_r\geq0$, $n(\xi)-2n(\varphi)=n_l\geq0$ and
$n_r+n_l-2n(\theta)\geq0$. In addition to these conditions 
$2-|\theta|\not=0$ should also be required for all $\theta$ (as 
this has been used in deriving (\ref{3.12}) 
(any $\Theta=\pm2$ should
be represented by a $\vartheta$ or $\varphi$)). Constructing the simplest
solutions of (\ref{3.12}) we have found that this requirement can be satisfied 
if $n(\theta)\leq\min(n_r,n_l)$. 

We interpret Eqs.\ (\ref{3.5}a-b), (\ref{3.8}a-b) and 
(\ref{3.9}a-b) as the secular equations of
the massive (`m') and the two (`r' and `l') kinds of massless particles, and 
we argue that Eq.\ (\ref{3.12}) describes how the two massless sectors combine.
In connection with their structure and content we note the following:
\begin{itemize}
\item[i.] Eqs.\ (\ref{3.5}a-b) are of the BA form: 
this structure is typical for the 
HLBA equations of systems with excitations having an internal degree of 
fredom. In the case of the massive particles this degree of freedom is the 
spin (see (\ref{2.12})). A solution of the secular equations corresponds 
to a state 
\bee
 S^2=l(l+1)\,,\quad S^z=m\,,\quad l=m={n(\kappa)-2n(\chi)\over2}\,,\label{3.13}
\ee
i.e.~there are $n(\kappa)$ particles, $n(\kappa)-n(\chi)$ with up and
$n(\chi)$ with down spins. States with $m<l$ can be constructed
from these by acting on them by the operator $S^-\propto(S^x-iS^y)$. If this 
operator is applied $\nu$ times, we still have $l=(n(\kappa)-2n(\chi))/2$
but $m=l-\nu$. 
This way, if (\ref{3.5}a-b) give all the $l=m$ states, 
all the $|m|\leq l$ states 
can be constructed. 
These states are degenerate with the corresponding
$l=m$ one. 
It should be noted, that the action of the $S^-$ operator can be 
implemented in the BA formalism 
by introducing as many extra $\chi$s as many 
times the $S^-$ acts and taking these extra $\chi$s to $+$ or $-\infty$. 

\item[ii.] (\ref{3.8}a-b), (\ref{3.9}a-b) and (\ref{3.12}) 
describe the massless sectors
where the internal degree of freedom is the isospin, with its third component
being the charge (particles with $I_3=\pm1/2$ have charges $\pm1$).  
Since the secular equations of the two massless (`l' and `r')
sectors are independent of each other (see item vi.~too) and have
the same structure as those of the massive particles, it is natural
to think that the isospins carried by the `l' and `r' particles (based on
the analogy to the spin of the massive sector) can be defined
separately. On the other hand
we can make exact statements only about the 
total isospin ((\ref{3.14b}), (\ref{3.15b}), (\ref{3.16b}) and 
(\ref{3.17b})), but not about the 
isospins of the `l' and 
`r' sectors separately. A reason for this is that in the original 
Hubbard model the `l' and `r' sectors (so their isospins) 
are not even defined, 
as the `l' and `r' sectors separate in the scaling limit only. 
In the following we try to resolve this apparent contradiction and propose an 
interpretation according 
to which it is possible to choose a basis, 
where both the `l' and `r' sectors are in isospin eigenstates decoupled from
each other.
\item[] A central element of our proposed interpretation is that in any state
characterized by $n_r$ and $n_l$ both massless sectors are in eigenstates of
the ${(\vec I)^2}$ with eigenvalues
\bee\label{inegyzk} 
\left(I^{(r)}\right)^2={n_r\over2}\left({n_r\over2}+1\right)\quad{\rm and}
\quad\left(I^{(l)}\right)^2={n_l\over2}\left({n_l\over2}+1\right).
\ee
This is supported by the following. 
If one of the sectors is empty (the `r' one, say), the other (the `l' sector)
carries the total isospin 
and its state is indeed an eigenstate of 
$(I^{(l)})^2$ and $I^{(l)}_3$, where the value of $(I^{(l)})^2$ is the one 
given above and $I^{(l)}_3=n_l/2$. Acting on this state by 
$I^{-}\!\propto I_1-iI_2$ decreases $I_3^{(l)}$ by one, leaving $(I^{(l)})^2$ 
(and all the other quantities) unchanged\footnote{As in the scaling 
limit $I^-$ may be not well defined, to apply $I^-$ means in fact to 
act on a state of the finite chain by $I^-$ first and take the scaling 
limit afterwards.}. 
By a repeated application of $I^-$, all the $\vert I_3^{(l)}\vert\leq n_l/2$ 
states 
can be constructed, i.e.~a single solution of 
Eq.~(\ref{3.9}a-b) represents a complete set of 
degenerate
states with $(I^{(l)})^2$ given above and $\vert I_3^{(l)}\vert\leq n_l/2$.
Analogous statements hold for the `r' sector if the `l' one is empty.
Based on the fact, that the solutions of (\ref{3.8}a-b) and 
(\ref{3.9}a-b)  
are {\em independent} of each other, 
it is very
plausible to assume that this also holds for the
states corresponding to them. Thus we propose that any state corresponding 
to a solution
of Eqs.~(\ref{3.8}a-b), (\ref{3.9}a-b) is built up as a combination of the
`r' and `l' states represented (in the above sense) by the given solutions
of Eqs.~(\ref{3.8}a-b) resp.~(\ref{3.9}a-b), 
even if none of the two sectors is empty.
\item[] The states with $n(\theta)=0$ satisfy,
in addition to Eqs.\ (\ref{inegyzk}), 
\bml
\bee
 I_3^{(r)}={n_r\over2},\quad I_3^{(l)}={n_l\over2}\,,
\ee
consistently with the exact formula for the total isospin,
\bee
I^2={n_r+n_l\over2}\left({n_r+n_l\over2}+1\right),\quad I_3={n_r+n_l\over2}\,.
\label{3.14b}
\ee
\eml
Acting on such a state by $I^-$ decreases $I_3$ by one,
leaving the value of $I^2$
unchanged.
The resulting state is a symmetric
linear combination of two states in which both $(I^{(r)})^2$ and
$(I^{(l)})^2$ are unchanged, but in one $I_3^{(r)}$, in the other $I_3^{(l)}$
is decreased by $1$.  
One can act with $I^-$ on a state several ($n_r+n_l$)
times.
If $I^-$ is applied $\nu$ times, the resulting state is given as 
such a linear combination of the states with
\bml\label{3.15}
\bee\label{3.15a}
  I_3^{(r)}={n_r\over2}-\nu_r\left(\geq-{n_r\over2}\right),\quad
  I_3^{(l)}={n_l\over2}-\nu_l\left(\geq-{n_l\over2}\right),\quad
  \nu_r+\nu_l=\nu\,,
\ee
for which the total isospin is
\bee
 I^2={n_r+n_l\over2}\left({n_r+n_l\over2}+1\right),
 \quad I_3={n_r+n_l-2\nu\over2}\,.
 \label{3.15b}
\ee
\eml
If $n(\theta)\not=0$, none of the sectors are in $I_3^{(r,l)}$ eigenstates, 
but such a state is built up as a suitable linear combination of states 
characterized by (\ref{inegyzk}) and
\bml\label{3.16}
\bee
  I_3^{(r)}={n_r\over2}-\nu_r\,,\quad 
  I_3^{(l)}={n_l\over2}-\nu_l\,,\quad
  \nu_r+\nu_l=n(\theta)\,,
  \label{3.16a}
\ee
so that the total isospin is
\bee
I^2={n_r+n_l-2n(\theta)\over2}\left({n_r+n_l-2n(\theta)\over2}+1\right),
\quad I_3={n_r+n_l-2n(\theta)\over2}\,,
\label{3.16b}
\ee
\eml
as required by (\ref{2.12}). (This interpretation is consistent with 
the following:
a $\theta$ can be considered as a $\vartheta\sim\infty$ or as a 
$\varphi\sim-\infty$, and (in analogy to what is said
in item i.) the presence of such infinite variables 
can be identified with 
the action of $(I^{(r)})^-$ resp.~$(I^{(l)})^-$.) 
Of course, if $n_r+n_l-2n(\theta)>0$, $I^-$ can 
also be applied to these
states further decreasing $I_3$ while leaving $I^2$ unchanged.  
This way, the action of
$I^-$ $\nu$ times ($n_r+n_l-2n(\theta)-2\nu\geq0$) results in a state, 
which is such a linear combination of the states with
\bml\label{3.17}
\bee
   I_3^{(r)}={n_r\over2}-\nu_r,\quad  
   I_3^{(l)}={n_l\over2}-\nu_l,\quad
   \nu_r+\nu_l=n(\theta)+\nu,\label{3.17a}
\ee
for which the total isospin is
\bee
I^2={n_r+n_l-2n(\theta)\over2}\left({n_r+n_l-2n(\theta)\over2}+1\right),
\quad I_3={n_r+n_l-2n(\theta)-2\nu\over2}.
\label{3.17b}
\ee
\eml
\item[] All the states with different $n(\theta)$ and $\nu$ corresponding 
to a given pair of solutions 
of Eqs.\ (\ref{3.8}a-b) and (\ref{3.9}a-b) are degenerate, but linearly
independent. 
Based on the above
interpretation, it is possible to choose such a basis
 in the space of these states,  
in which the basis vectors are labelled by the quantities $(I^{(i)})^2$,
$I^{(i)}_3$, $(i=l,r)$ (see also Appendix
\ref{sec:A}). We also note
that in this basis the length of the total isospin 
$(\vec I^{(r)}+\vec I^{(l)})^2$ is not a good quantum number any more.

\item[iii.] The chiral charge is given by
$C=2(I_3^{(r)}-I_3^{(l)})$. For a large 
class of BA eigenstates (described by (\ref{3.15}), (\ref{3.16}) and 
(\ref{3.17})) 
the chiral charge is not well defined, as these states are not eigenstates
of $C$. This is a consequence of the fact, that in the lattice model
the chiral charge is not even defined. The new basis vectors described 
in item {ii.}\ are, however, not only eigenstates of the energy and momentum, 
but also of the isospins of the two sectors {\em separately}.
For this reason they are 
eigenstates of the chiral charge too, ie.\ the spectrum is 
{\em chiral invariant}. The possible eigenvalues of the
chiral charge $C$  
for states characterised by $n_r$ and $n_l$ are 
\bee
C=n_r+n_l-2n\ ,\quad n=0,1,2,\cdots,n_r+n_l\ .\label{3.18}
\ee

\item[iv.] The scattering phaseshifts of the physical particles can be 
reconstructed from their secular equations. The method for this is based
on the idea, that the deviations of the particle momenta from the free 
values can be interpreted as the phaseshifts of the particles scattering
on each other \cite{AnLo2,Kor,DeLo1}. Consider a two particle 
scattering-state on a ring. If the momenta are $p_1$ and $p_2$ then   
periodic boundary condition requires $Lp_1+\delta_{12}=2\pi n_1$ and
$Lp_2-\delta_{12}=2\pi n_2$ with $n_1$ and $n_2$ integers and $\delta_{12}$
being the phaseshift. Writing the scaling limits of the HLBA equations
in this form the phaseshifts can be found. 
\item[] A triplet state of two massive particles is described by two  
$\kappa$s and no $\chi$s. For such a state Eq.\ (\ref{3.5a}) yields
\bee\label{deltatr}
\delta^{tr}_{12}=\phi\left({x\over\pi}\right)+\pi\,,
\quad x=\kappa_1-\kappa_2\,,
\ee
where the $\pi$ comes from the parity prescription for the parameters $I_j$.
A singlet state is characterised in addition to $\kappa_1$ and $\kappa_2$ 
by a $\chi$ for which Eq.\ (\ref{3.5b}) yields $\chi=(\kappa_1+\kappa_2)/2$. 
For this case the above reasoning leads to 
\bee\label{deltasing}
\delta^{s}_{12}=\phi\left({x\over\pi}\right)-
2\tan^{-1}\left({x\over\pi}\right)\,.
\ee
>From the phaseshifts (\ref{deltatr}) and (\ref{deltasing})
the $S$-matrix of an `m-m' scattering can be given as 
\bee\label{smm}
\hat S_{mm}(x)=-\exp\left\{i\phi\left({x\over\pi}\right)\right\}
{x{\hat{\rm I}}-i\pi{\hat \Pi}\over x-i\pi}\,,
\ee
where $\hat{\rm I}$ and $\hat \Pi$ are the identity resp.\ permutation operators 
acting on the spins of the two particles:
\bee
{\rm I}^{\sigma'_1\sigma'_2}_{\sigma_1\sigma_2}=
\delta_{\sigma'_1\sigma^{\phantom{'}}_1}
\delta_{\sigma'_2\sigma^{\phantom{'}}_2}\ ,\quad
\Pi^{\sigma'_1\sigma'_2}_{\sigma_1\sigma_2}=
\delta_{\sigma'_1\sigma^{\phantom{'}}_2}
\delta_{\sigma'_2\sigma^{\phantom{'}}_1}\ .
\ee
In a similar way the $S$-matrices 
for the `r-r' and `l-l' scatterings are found to be 
\bee\label{slr}
\hat S_{rr}(x)=\hat S_{ll}(x)=-\hat S_{mm}(x)
\ee
with $\hat{\rm I}$ and $\hat\Pi$ acting on the isospins, and 
$x=\eta_1-\eta_2$ resp.\ $x=\xi_1-\xi_2$ for the `r' resp.\ `l'
sector.  
The scattering matrices (\ref{smm}), (\ref{slr}) coincide
-- up to an overall sign -- with those 
given in Ref.\ \cite{Me}. (Actually it is argued in Ref.~\cite{EssKo} that
in general
$S$-matrices can be determined only up to an overall phase
from the HLBA equations, 
and  further considerations are needed to determine this phase ambiguity.)

\item[v.] The `m-r', `m-l' and `r-l' phaseshifts do not appear 
in the equations explicitly.
Since they are just constants, we absorbe them into the 
quantum numbers $I_j$, $J_f$ and $J_q$. The prescriptions for these
quantum numbers are compatible with these phaseshifts being $+$ or 
$-\pi/2$ as obtained in Ref.\ \cite{Me}. 

\item[vi.] We note, that the `r' and `l'  massless 
sectors are apparently independent of each other and the massive sector
in the sense that in the Eqs.\ (\ref{3.8}a-b) and (\ref{3.9}a-b)
determining the rapidities of these sectors no parameters of the
other sectors appear, i.e.\ the rapidities $\eta$ and $\xi$ are independent 
of the states of the other sectors.
This is a consequence of the requirement that the total number 
of excitations should be even (Eq.\ (\ref{2.11})). To see this consider a 
left moving (`l') particle. In the equation determining its rapidity 
(besides the phaseshifts due to scattering on the other `l' particles)
the sum of the phaseshifts corresponding to scattering on the massive and
`r' particles appear.
This sum can be expressed (mod 2$\pi$) 
by the number of the `l' particles only, and this
is why no parameters referring to the `m' and `r' sectors appear in 
(\ref{3.9}a-b). Similar statements hold for the `r' sector, but
{\em not} for the equations of the massive particles.
The phaseshifts of the `m-l' and `m-r' scattering  
have an opposite sign, so it is
the difference of the number of the `l' and `r' particles which enters into 
$I_j$ of (\ref{3.5a}). This suggests that in spite of the apparent 
independence
of the massless sectors the different sectors are nontrivially coupled.
(This is also indicated by the prescription that the total number 
of excitations should be {\em even}.)
\end{itemize}

\section{CONFORMAL PROPERTIES}
\label{sec:conform}

The existence of a massless sector signals an underlying 
conformal field theory. By calculating the finite
size corrections to the ground-state energy and the fine structure
of the spectrum 
arising due to the quantization of the momenta
(tower structure) one finds the central charge and the
conformal weights \cite{Car}.
To find the 
finite size corrections to the ground-state (dressed vacuum) energy 
to the spectrum one has to calculate them for the Hubbard chain and take the 
scaling limit afterwards. 
In this Section we give first the finite
size corrections to the spectrum of Eq.~(\ref{2.9a}), then we take the 
limit (\ref{1.3})
and finally we show that the resulting spectrum has the conformal tower 
structure. (An alternative treatment is to consider the lowest
energy states in the finite $N$ Hubbard chain, and take the scaling
limit of these states directly (as it is described in Appendix 
\ref{sec:vegesm}). Both treatments we present are based on an Euler-Maclaurin
type summation formula, but there exist other methods too \cite{KlPe,PeKl}.) 

When deriving equations (\ref{2.7}a-b) and (\ref{2.8}a-b) and
the spectrum   
(\ref{2.9}a-b) from Eqs.\ (\ref{2.4}a-b) and (\ref{2.5}), the summations 
over the
$\Lambda$s have been replaced by integrals: 
$(1/N)\sum_{\eta}f(\Lambda_{\eta})\to\int f(\Lambda)\sigma(\Lambda) d\Lambda$   
with $\sigma(\Lambda)$ being the density of the $\Lambda$.
This replacement becomes exact in the $N\to\infty$ limit only. For large
but finite $N$ using an Euler-Maclaurin type formula even the 
$(1/N)\sum_{\eta}f(\Lambda_{\eta})-\int f(\Lambda)\sigma(\Lambda) d\Lambda$
difference can be taken into account. The corrections obtained this way 
are referred to as
finite size corrections. The method is elaborated to quite an extent
\cite{Wo4},
and it can be 
adapted for the present treatment (details will appear elsewhere). 
As a result (\ref{2.9a}) gets modified, an additional term appears:
\bee
E=N\varepsilon_0-{\pi tI_1(\pi/2u)\over3I_0(\pi/2u)N}
  +\sum_j \varepsilon_s(k_j)+\sum_h\bar\varepsilon_c(\Lambda_h)\,,
  \label{4.1}
\ee
and $\bar\varepsilon_c(\Lambda_h)$ differs from 
$\varepsilon_c(\Lambda_h)$ by terms $O(1/N)$.
In the limit (\ref{1.3})  $t$ can be replaced by $1/2a$, and after
taking $u\to0$ we have
\bee
\ba{c}
{\displaystyle E-N\varepsilon_0=-{\pi\over6L}+\sum\epsilon_m(\kappa)+
  \sum\bar\epsilon_r(\eta)+\sum\bar\epsilon_l(\xi)}\ ,\\
{\displaystyle  P=\sum^{\vphantom{l}} p_m(\kappa)+\sum\bar p_r(\eta)+ 
\sum\bar p_l(\xi)}\ .
\ea\label{4.2}
\ee
For energies of the order of unity $\bar\epsilon_{r,l}=\epsilon_{r,l}$
(and $\bar p_{r,l}=p_{r,l}$), but for small energies (of the order of $1/L$) 
the form of these quantities slightly deviates from those valid for larger
energies. It is important to note, that
\bee
\bar\epsilon_{r}=\bar p_{r}\quad{\rm and}\quad\bar\epsilon_{l}=-\bar p_{l}
\ee
still hold. The secular equations for the massive particles is not 
effected by the 
finite size corrections, but those for the massles particles are. For the
simplest case (when there are no $\vartheta$s and $\varphi$s, these states are
of the lowest energies within a tower) they read
\bee\ba{c}
 {\displaystyle L\bar p_r(\eta_f)=2\pi J_f-
 \sum_{f'}\bar\phi\left(\eta_f,\eta_{f'}\right)\,,}\\
 {\displaystyle J_f=-\left({n_r\over4}+{1\over2}\right)\modegy\,,}
 \ea\label{jobbra}
\ee
and
\bee\ba{c}
 {\displaystyle L\bar p_l(\xi_q)=2\pi J_q-
 \sum_{q'}\bar\phi\left(\xi_q,\xi_{q'}\right)}\,,\\
 {\displaystyle J_q=\left({n_l\over4}+{1\over2}\right)\modegy}\,.
 \ea\label{balra}
\ee
Also here, for energies of the order of unity 
$\bar\phi\left(x,x'\right)=\phi\left(x,x'\right)$, but for
small energies the two quantities deviate (but even for small energies
$\bar\phi\left(x,x'\right)=-\bar\phi\left(x',x\right)$).
We note, that the structure of these equations, and that of the energy
and momentum are the same as they would be without the finite size
corrections, and to calculate the spectrum of the simple states we do not 
even need the actual form of the $\bar\epsilon_{r,l}$, $\bar p_{r,l}$ and
$\bar\phi$.  

If there are no excitations, 
on the right hand side of the first equation in (\ref{4.2}) only the 
first term  remains,
showing that the universal part of the ground state energy
of the massless sector is the same as 
that of a Gaussian field with conformal anomaly $c=1$.

Next we argue that the contribution of the 
massless excitations in (\ref{4.2}) 
has the conformal tower structure, provided the variables 
$\eta$ and $\xi$
satisfy Eqs.\ (\ref{jobbra}) and (\ref{balra}). By summing 
these equations we get 
\bee
\ba{l}
  {\displaystyle \sum\bar\epsilon_r(\eta)+\sum\bar\epsilon_l(\xi)=
  {2\pi\over L}(\J_r+\J_l)}\ ,\\
  {\displaystyle \sum^{\vphantom{N}}\bar p_r(\eta)+ \sum\bar p_l(\xi)=
  {2\pi\over L}(\J_r-\J_l)}\ ,
\ea\label{4.3}
\ee
with 
\bee\label{4.4}
 \J_r=\sum J_f\,,\qquad
 \J_l=-\sum J_q\,.
\ee
It is not hard to see (based on the prescriptions for the $J$
quantum  numbers), that         
\bee\label{4.5}
 {\displaystyle \J_r=-{n_r(n_r+2)\over4}\modegy}\,,\qquad
 {\displaystyle \J_l=-{n_l(n_l+2)\over4}\modegy}\,.
\ee
Consequently the states with given
$n_r$
and $n_l$ {\em form towers\/}: 
at a fixed $n_r$ all possible $\J_r(n_r)$ differ by integers, 
i.e.\ $\J_r(n_r)=\J_{r(min)}(n_r)+N_r$ and similar statements hold for
$\J_l(n_l)$. Thus 
\bee
\ba{l}
 {\displaystyle {2\pi\over L_{\vphantom{j}}}(\J_r+\J_l)={2\pi\over L}
  (\J_{r(min)}(n_r)+\J_{l(min)}(n_l)+N_r+N_l)}\ ,\\
 {\displaystyle {2^{\vphantom{l}}\pi\over L}(\J_r-\J_l)=
  {2\pi\over L}(\J_{r(min)}(n_r)-\J_{l(min)}(n_l)+N_r-N_l)}\ .
\ea\label{4.6}
\ee 
$\J_r$ takes its smallest value, if 
\bee
J_f:\quad-{n_r\over4}+{1\over2},\ \ -{n_r\over4}+{3\over2},\ \cdots,\ 
-{n_r\over4}+{2n_r-1\over2}\,,\label{4.7}
\ee
as if $J_{f(min)}<-n_r/4+1/2$ Eq.~(\ref{jobbra}) has no solution.
Then
\bml\label{4.8}
\bee
\J_{r(min)}(n_r)={n_r^2\over4}\ .\label{4.8a}
\ee
In a similar way we have
\bee
\J_{l(min)}(n_l)={n_l^2\over4}\ ,\label{4.8b}
\ee
\eml
thus for a given $n_r$ and $n_l$ the minimal energy $\Delta E$ and the 
corresponding momentum $\Delta P$ of the massless excitations (the apex
of the $\{n_r,n_l\}$ tower) are
\bee
  \Delta E=
  {\displaystyle {2\pi\over L}\left({n_r^2\over4}+{n_l^2\over4}\right)},\quad
  \Delta P=
  {\displaystyle {2\pi\over L}\left({n_r^2\over4}-{n_l^2\over4}\right)}.
\label{4.9}
\ee
According to these the conformal weights are 
\bee
\Delta={n_r^2\over4},\quad\bar\Delta={n_l^2\over4},\label{4.10}
\ee
i.e.\ the corresponding scaling dimensions and spins are given as
\bee\label{xs}
x={n_r^2\over4}+{n_l^2\over4},\quad s={n_r^2\over4}-{n_l^2\over4}.
\ee
(Here we use the notations of Cardy \cite{Car}.)
Introducing 
\bee\label{nm} 
n={n_r+n_l\over2}\,,\quad m={n_r-n_l\over2}\,.
\ee
Eqs.\ (\ref{xs}) takes the form
\bee\label{xsnm}
x={1\over2}\left(n^2+m^2\right)\,,\quad s=nm\,.
\ee
This structure formally coincides with the expected one for a $c=1$ 
conformal field theory coresponding to an SU(2) WZW theory of
level one, however,
this correspondence is not complete since $n$ and $m$ can also take
half-integer values!
When the number of massive particles is {\em even} both $n$ and
$m$ are integers, while
if the number of massive particles is {\em odd}, $n_r\pm n_l$ is
odd, and both 
$n$ and $m$ are half-integers. (This is a consequence of the requirement, 
that the total number of excitations must be even, as given by
(\ref{2.11}) and is
also discussed in item vi.\ of the previous section.)
Thus we can interpret (\ref{xsnm}) as follows: when the massive sector is 
empty
the massless sector can be described by an SU(2) WZW theory 
of level one. When the massive sector is not empty
one cannot expect a simple conformal field theoretic description
because of the coupling of the massless 
particles to the massive ones 
(as discussed in items v.\ and vi.\ of the previous section).
Also in this case the 
energy contribution of the massless sector shows a tower structure, but the 
interpretation of the apexes of the towers as scaling dimensions and spins
is not straightforward. 

It is to be noted, that in the states with no $\theta$s the parameters $n$
and $m$ can be identified as $n=I_3$ and $m=C/2$ with $C$ being the chiral
charge. For states with a finite number of $\theta$s such an interpretation
is not possible.
\goodbreak
\section{STATES OF FINITE DENSITIES}
\label{sec:dens}

Although in our derivation it has been
supposed, that the number of excitations is small, the resulting equations 
(\ref{3.5}), (\ref{3.8}) and (\ref{3.9})
can be used to describe states with a finite density of excitations too:
even if the number of excitations is macroscopic, i.e.\ $\sim\rho L$, 
that is small compared 
to $N$ since $\rho L/N=\rho a\to0$ in the scaling limit. In this section 
we discuss the lowest energy finite density states. As the different sectors 
are `almost decoupled' (Sec.\ \ref{sec:rellim}.), 
this can be done for the different sectors separately.
(The coupling made explicit by the parity-prescription for $I_j$ can give a 
contribution $O(1/L)$ in the energy density.)

It is plausible to assume that
for the lowest energy state of the massive sector (at a fixed number of 
particles) the 
$I_j$ quantum numbers are consecutive integers or half-odd-integers centered
around the origin.
If in such a 
state the number of $\kappa$s is macroscopic ($N_m$), they can
be described by a 
density $\sigma(\kappa)$ determined from Eq.\ (\ref{3.5a}) as 
\bee\label{sumsigma}
\sigma(\kappa)={1\over2\pi}{d\over d\kappa}\left(
p_m(\kappa)-{1\over L}\sum_{\alpha}2\tan^{-1}
  {\kappa-\chi_{\alpha}\over\pi/2}+{1\over L}
  \sum_{j'}\phi\left({\kappa-\kappa_{j'}\over\pi}\right)\right)
\ee
It is then clear, that the density will be the highest possible around the 
small $\kappa$s (small energies) if there are no $\chi$s. (Strictly speaking 
this argument holds for real $\chi$s only, but it can be modified to hold for
more complicated $\chi$ configurations too.) After replacing
the sum over the $\kappa_{j'}$s by an integral we find that $\sigma(\kappa)$
satisfys the equation
\bee\label{intenergy}
\sigma(\kappa)-\int\limits_{-B}^{B}d\kappa'K(\kappa-\kappa')\,
\sigma(\kappa')={m_0\over2\pi}\cosh\kappa\,,
\ee
where the kernel $K(\kappa)$ can be written in terms of the $\phi$ of 
(\ref{1.5}) as
\bee
K(\kappa)={1\over2\pi}{d\over d\kappa}\phi(\kappa/\pi)\,.
\ee
The $\sigma(\kappa)$ is also subject to the condition   
\bee\label{intenergy1}
\int\limits_{-B}^{B}d\kappa\,\sigma(\kappa)={N_m\over L}=\rho\,.
\ee 
In terms of the solution of Eq.~(\ref{intenergy}) the energy density
${\hat{\cal E}}(\rho)$ is given as
\bee
{\hat{\cal E}}(\rho)=m_0\int\limits_{-B}^{B}d\kappa\,\sigma(\kappa)\cosh\kappa\,.
\ee 
(Equations (\ref{intenergy}-\ref{intenergy1}) are exactly the
same as the ones obtained in Ref.~\cite{DeLo2} 
directly from the BA equations of the CGN model.)
We present the explicit solution
of Eqs.\ (\ref{intenergy}-\ref{intenergy1}) for a large density
($\rho/m_0\gg1$) using the results of Refs.~\cite{FoNi,FoNiWe}.
To apply those
results directly we also give the integral
equation determining the zero temperature free energy, obtained from Eqs.\ 
(\ref{intenergy}) and (\ref{intenergy1}) by a Legendre transformation:
\bee\label{free}
\hat f(\mu):=\mathrel{\mathop{\rm min}_{\displaystyle\rho}}
({\hat{\cal E}}(\rho)-\mu\rho)\,,
\ee
where $\mu$ denotes the chemical potential. After a straightforward 
manipulation one finds that 
\bee
\hat f(\mu)=-{m_0\over2\pi}\int\limits_{-B}^{B}d\kappa\,\epsilon(\kappa)\cosh\kappa\,,
\ee  
where the function $\epsilon(\kappa)$  and the `Fermi point', $B$,
are determined by the equation
\bee\label{inteq}
\epsilon(\kappa)-\int\limits_{-B}^{B}d\kappa'K(\kappa-\kappa')
\,\epsilon(\kappa')=\mu-m_0\cosh\kappa\,,
\ee
together with the condition $\epsilon(\pm B)=0$.
It is now a simple matter to apply the results of
Ref.~\cite{FoNi,FoNiWe} to find the solution for a large density
of Eqs.\ (\ref{inteq}). 
The asymptotic expansion of free energy is found to be given as:
\bee \label{fba}
\hat f(\mu)=-{\mu^2 \over2\pi}\cdot2  
\left\lb 1-{1\over2 \ln(\mu/m_0)}-
{\ln \ln(\mu/m_0)\over4 \ln^2(\mu/m_0)}-
{A\over8\ln^2(\mu/m_0)}+O({\ln \ln(\mu/m_0)\over\ln^3(\mu/m_0)})\right\rb\,,
\ee
where the constant $A$ is
\bee \label{cn}
A= 1+2\ln\pi -6\ln2 \, .
\ee
The energy density of the system is easily derived from 
(\ref{fba}) using the definition of the free energy
with the result
\bee \label{erho}
{\hat {\cal E}}(\rho) ={\pi \rho^2 \over 4}
\left\lb 1+{1\over 2\ln (\rho/m_0)}
+ {\ln\ln(\rho/m_0) \over 4\ln^2 (\rho/m_0)}
+{3-2\ln(2\pi) \over8 \ln^2(\rho/m_0)} +O({\ln\ln(\rho/m_0) \over\ln^3(\rho/m_0)}
\right\rb\,.
\ee
The leading term agrees with that of Ref.~\cite{DeLo2},
the non leading terms disagree, however.

The treatment of the massless sector due to the linear dispersion 
is much simpler than that of the massive one as we need not 
solve an integral equation. Consider for definiteness
the `r' sector (the treatment of the `l' sector is completely analogous). 
In the lowest energy states there are
no $\vartheta$s, so if the number of particles 
is $N_r=\varrho_r L$ the $J_f$ 
quantum numbers are (just as in (\ref{4.7}))   
\bee
J_f:\quad-{N_r\over4}+{1\over2},\ \ -{N_r\over4}+{3\over2},\ \cdots,\ 
-{N_r\over4}+{2N_r-1\over2}
\ee
leading to the energy density of the `r' sector
${\cal E}_r(\varrho_r)={\pi\varrho_r^2/2}$
and if $\varrho_r=\varrho_l=\varrho/2$, the total energy density
of the massless sector is
\bee
{\bar{\cal E}}(\varrho)={\pi\varrho^2\over4}\,.
\ee
The free energy density of the massless sector (defined as
the Legendre transformation of ${\bar{\cal E}}(\varrho)$ with
a chemical potential $\nu$) is   
\bee\label{fre}
\bar f(\nu)=-{\nu^2\over\pi}\,.
\ee

We note here that the above results can also be obtained 
directly from the Hubbard chain, by constructing a state
with a finite density of {\em excitations in the presence of 
chemical potential and magnetic field} 
and take the relativistic limit afterwards.

\goodbreak
\section{RELATION TO THE CHIRAL GROSS--NEVEU MODEL}
\label{sec:cgn}

In this section we present some evidence supporting
the equivalence of the theory defined by the
scaling limit of the Hubbard model
with the SU(2) CGN model. In fact, based on the perturbative and 
non perturbative results it seems to as that the 
the full renormalized CGN theory can be {\em defined} as
the scaling limit of the HF Hubbard chain.
 
The Lagrangian of the SU(n) symmetric CGN model \cite{GrNe} can be
written as:
\bee \label{lagr0}
{\cal L} =  i\bar{\psi_a} \delsl \psi_a
+ {1 \over 2} g_s \left\lb (\bar{\psi_a} \psi_a)^2
  +(\bar{\psi_a} \gamma_5 \psi_a)^2 \right\rb
- {1 \over 2} g_v  (\bar{\psi_a} \gamma_{\mu} \psi_a)^2 \,,
\ee
where the $\psi_a$ are Dirac fermions, $a=1\,,\ldots n$ and summation
over the repeated $a$ indices is understood. 
The theory given by the Lagrangian (\ref{lagr0}) 
is invariant under
a global U($n$)$\otimes$U$(1)_{\rm chiral}$ transformation;
\bee\label{symtr}
\psi'_a=M_{ab}\psi_b\,,\qquad \psi_a''=e^{\alpha\gamma_5}\psi_a\,,
\ee
with $M\in$U($n$) and $\alpha$ being a (real) constant. (We note that
the couplings ($g$, $g'$) in (\ref{lagr1}) are related to those of
(\ref{lagr0}) as $g=g_s$, $g'=g_v+g_s/2$.)

As mentioned  already in the Introduction the CGN model is asymptotically
free \cite{MiWe}, and as first found in the large $n$ expansion  
its spectrum contains massive particles \cite{GrNe}.
Furthermore due to  
an infinite number of conservation laws preventing 
particle production, 
the $S$--matrix  of the CGN model is factorizable, i.e.\ the multiparticle $S$--matrix
can be written as a product of the two--particle $S$--matrices.
Using the `bootstrap' approach (when the spectrum of the
theory is postulated) the 
$S$--matrix of the model had been proposed long time
ago \cite{BKKW,ABW,KoKuSw}. From now on we concentrate on the SU(2)
symmetric CGN model, whose massive
spectrum is particularly simple as it consists of
a single doublet.   
Note that in this case the particles are their own antiparticles.

The bootstrap approach led to the following two--particle $S$--matrix
of the massive  doublet in the symmetric channel \cite{ABW}:
\bee \label{smat}
 S(x)=
{\Gamma(1+i{x/2\pi})\Gamma(1/2-i{x/2\pi})\over
\Gamma(1-i{x/2\pi})\Gamma(1/2+i{x/2\pi})}~,
\ee
where $x$ denotes the rapidity of the particles. (Note that $S(x)$ coincides
with $\exp\{i\phi(x/\pi)\}$ where $\phi(x)$ is the `phaseshift'-function 
in Eq.~(\ref{1.5}).)

As it has been argued in Ref.~\cite{Wit} using bosonization 
there is no contradiction between the
existence of massive particles and the necessarily unbroken
(continuous) chiral symmetry of the theory. 
The following picture of the CGN model has by now emerged \cite{AnLo1,MoScha}:
the theory decouples into a massless and a massive sector. 
The massless excitations carry a U($1$) charge but
are singlets under SU(2).
The massive sector consists of particles with a dynamically generated mass
transforming non--trivially under SU($2$) but without a U($1$) charge.

The application of the BA to 
the Hamiltonian of (\ref{lagr0}) \cite{AnLo1,Bel}  was a very important 
development as it made possible to obtain exact, nonperturbative results.
In Ref.~\cite{AnLo1} the theory has been regularized by filling the Dirac sea
up to a certain depth $K$ (in a finite volume). This regularization
is rather unorthodox as compared to other schemes,
and
it would be clearly desirable to work with a more physical cutoff.
The main problem with other cutoff
schemes is that they spoil integrability in general. Therefore it appears 
to be of particular interest if one could find an {\em integrable}
lattice regularization of the CGN model. 
At first sight it is not immediately obvious that 
the Hubbard model would be a good candidate for a lattice version
of the CGN model. The naive continuum limit ($a\rightarrow0$) 
of the half filled
Hubbard model is given by the following Lagrangian (See Appendix 
\ref{sec:nlim})
\bee \label{naivlim}
{\cal L} =  i\bar{\psi_a} \delsl \psi_a
-u\left\lb (\bar{\psi_a} \psi_a)^2
   +(\bar{\psi_a} \gamma_{0} \psi_a)^2\right\rb \,,
\ee
which significantly differs from (\ref{lagr0}), as the theory
corresponding to (\ref{naivlim}) lacks chiral invariance and it is not even
Lorentz invariant. The theory defined by (\ref{naivlim}) is in fact 
rather similar to the (nonchiral) Gross-Neveu model. 
The lack of chiral invariance in the naive limit is
not surprising as in the half filled case the umklapp processes play
an important role. 
Nevertheless there is a large body of evidence
that the scaling limit of the half filled Hubbard model
is actually the SU(2) CGN model. 
We list some of them here and then come back to the problem of the
naive continuum limit.

\begin{itemize} 
\item The present analysis of the HLBA  reproduces the expected spectrum 
of the CGN model
consisting of a massive doublet and two (`r' and `l') massless doublets.
The calculated phaseshifts are in accordance with the 
previously obtained results for the massive sector
\cite{ABW,AnLo2} and with those 
of Ref.~\cite{Me} where the complete $S$ matrix (including the
massless `r' and `l' doublets) has been found. The results show 
that the massive and massless sectors
are practically (but {\em not completely}) decoupled.
The massless sector is far from being trivial as
the left- resp.\ right-movers scatter on each other exactly in the same way
as the massless scattering theory proposed for a level one
WZW theory \cite{Zam2}. The $S$ matrix shows an 
SU(2)$\times$SU(2) (in fact SO(4)) symmetry which is in fact to
be expected
as this is a symmetry of the Hubbard model which is preserved in the
scaling limit. It is less clear, how
such an SO(4) symmetric $S$ matrix and spectrum can correspond to an SU(2) 
symmetric model.
In fact this discrepancy is removed by the 
observation that the SU(2) CGN model also possesses
a (hidden) SO(4) symmetry.
To exhibit this hidden SO(4) symmetry we
introduce four Majorana fermions $\chi_i$ ($i=1\,\ldots4$)
instead of the complex SU(2) doublet $\psi_a$ ($a=1\,,2$)
$\psi_1=\chi_1+i\chi_2$, $\psi_2=\chi_3+i\chi_4$. Then 
using the Majorana properties together with the Fierz identities 
it is easy to show that there are two inequivalent SO(4) symmetric
four fermion interactions in two dimensions: 
\bee
{\cal L}_i^{(0)}=(\bar\chi_i\chi_i)^2 \qquad
{\cal L}_i^{(1)}=
\epsilon^{ijkl}\bar\chi_i\gamma_5\chi_j\bar\chi_k\gamma_5\chi_l\,.
\ee
In terms of Dirac fermions these interactions can be written as:
\bee\label{inv}
{\cal L}_i^{(0)}=(\bar\psi_a\psi_a)^2\,,\qquad
{\cal L}_i^{(1)}=-2\left[(\bar\psi_a\gamma_5\psi_a)^2+
\h(\bar\psi_a\gamma_\mu\psi_a)^2\right]\,.
\ee 
Comparing (\ref{lagr0}) and 
(\ref{inv}) one can see, that  the SU(2) CGN 
model with $g_v+g_s/2=0$ possesses a hidden SO(4) symmetry
(in analogy to the nonchiral SU(2) 
Gross-Neveu model 
whose full symmetry is in fact O(4)). 
It is worth mentioning at this point that on the other hand
in the scaling limit of
HF Hubbard chain, eigenstates of the chiral charge can be defined
(i.e.~in that limit the chiral symmetry {\em not present} in the lattice 
model, appears). 
\item The detailed analysis of the finite size corrections
in section IV shows in a rather convincing way that the massless sector
corresponds to a $c=1$ CFT  precisely at the SU(2)$\times$SU(2)
symmetry point, corresponding to a level one WZW theory.  
\item  The calculation of the free energy for macroscopic densities
shows again the decoupling of the massive and massless sectors.
More importantly as the free energy is a physical quantity it can be 
directly compared with the results of a {\em perturbative} calculation
in the CGN model \cite{FoNi} valid for high densities because of asymptotic
freedom of the model. 
Using the results of Ref.~\cite{FoNi} 
the free energy calculated in perturbation theory can
be easily shown to be in complete agreement with that of the
HLBA equations (\ref{fba}), (\ref{fre}).
In Appendix \ref{sec:fren} we
outline the strategy of the calculation of the perturbative
free energy and simply state the result.

Comparing the 
free energy (\ref{fba}) obtained from the HLBA equations 
with the perturbative one (\ref{fp2}) one has to express
the chemical potential $\mu$ in terms of $h$ and $Q$ in Eq.~(\ref{fba}).
Since $\mu=hQ/2$ it is easy to verify that the perturbative result
(\ref{fp2}) is in perfect agreement with the exact one (\ref{fba})
provided 
\bee\label{mpla}
{m_0\over\Lambda_{\scriptscriptstyle\MSb}}={1\over\sqrt{2\pi}}\,.
\ee 
\item The beta-function of the coupling $u$ of the 
theory defined by the scaling limit of the HLBA system of the half filled
Hubbard model
can be immediately read off from Eq.~(\ref{1.3}):
\bee
\beta^{(u)}:=\mu{d\over d\mu}u=-\beta^{(u)}_0u^2-\beta^{(u)}_1u^3+
\ldots\qquad\mu={1\over a}\,,
\ee
where
\bee\label{hubbf}
\beta^{(u)}_0={2\over\pi}\,,\quad\beta^{(u)}_1=-{2\over\pi^2}\,.
\ee
Comparing Eqs.~(\ref{cgnbf}) and (\ref{hubbf}) one sees at once that the
beta-functions coincide up to two loops if one could identify
$u$ with $g$. On the basis of the naive continuum limit (\ref{naivlim})
this is of course not possible as it would rather suggest $u\propto g/2$
corresponding to the (nonchiral) GN model.
The perturbative
free energy  Eq.~(\ref{fp2}) contains, however, the 
two loop beta-function of the CGN model already as an input, and 
since Eq.~(\ref{fp2}) is in {\em perfect agreement}
with the result obtained in the scaling limit of the Hubbard model
(\ref{fba}) 
this provides a rather strong (though indirect) evidence that the two loop
beta-functions do in fact agree. It would be clearly desirable, however, to
show in a more direct way the equivalence
of the beta-functions and understand to some degree
the problems
encountered after having taken the naive limit (\ref{naivlim}).
\end{itemize}
The one-loop renormalization group analysis outlined in Appendix
\ref{sec:coupl} gives an indication that the rather pathological
theory obtained in the {\em naive} limit (\ref{naivlim}) would eventually
flow to the CGN model. (We note that this result has already been indicated 
by the ``g-ology'' treatment of the 1D electron gas \cite{So}.) As the
renorm trajectory of (\ref{naivlim}) approaches that of the CGN model
the couplings become large and the one loop analysis breaks down and that
is why the one loop analysis is not conclusive.

\acknowledgments
We are grateful to Profs.~N.~Andrei and P.~Weisz for drawing our attention
to some of the 
problems analysed here, and also to Prof.~J.~S\'olyom for discussions.
Supports from OTKA under grants numbers T014443 and T016233 
are acknowledged.

\appendix

\section{}\label{sec:A}

In this Appendix we would like illustrate our proposition about the isospin 
structure of the limiting states on an explicit example.

Suppose we have a certain $(\eta,\vartheta)$ set with $n_r=3$ solving 
Eqs.\ (\ref{3.8}a-b) 
and a $(\xi,\varphi)$ set with $n_l=2$
satisfying Eqs.\ (\ref{3.9}a-b). In the following we enumerate the possible 
isospin-structures of states characterized by such a set $(\eta,\vartheta)$
and $(\xi,\varphi)$. 
For the considered values of $n_r$ and
$n_l$ there are
three possibilities: there are no $\theta$s, there is one $\theta$ 
(which has the value $\theta=-2/5$ due to (\ref{3.12})), 
and there are two $\theta$s
(whith values $\theta^{\pm}=-2(1\pm i\sqrt{2})/3$). This way the  
the BA eigenstates (longest projection states) are
\bee\ba{l}
 \Big\vert\eta,\vartheta;\xi,\varphi;{5\over2},{5\over2}\Big\rangle\ ,
  \qquad n(\theta)=0\ ;\\
 \Big\vert\eta,\vartheta;\xi,\varphi;{3\over2},{3\over2}\Big\rangle\ ,
   \qquad n(\theta)=1\ ;\\
 \Big\vert\eta,\vartheta;\xi,\varphi;{1\over2},{1\over2}\Big\rangle\ ,
   \qquad n(\theta)=2\ .\ea
     \label{A1}
\ee
(Here the two numbers refer to the length and the third component of the 
isospin.) As we argued, based on the independence of 
the `r' and `l' sectors it is natural
to think that if $n(\theta)=0$ both sectors are longest projection isospin
eigenstates and the state {\em factorizes}:
\bee
  \Big\vert\eta,\vartheta;\xi,\varphi;
  {\textstyle{5\over2}},
  {\textstyle{5\over2}}\Big\rangle=
  \Big\vert\eta,\vartheta;
  {\textstyle{3\over2}},
  {\textstyle{3\over2}}\Big\rangle_r
  \Big\vert\xi,\varphi;{1},{1}\Big\rangle_l\ .\label{A.2}
\ee
(Although this is a natural assumption, it is not trivial: before taking 
the scaling limit the two sectors are not separated, so they and their
isospins are not defined;
while in the scaling limit the wavefunction does not exists, so (\ref{A.2}) 
can not
be directly checked. Nevertheless the assumption, that the isospins of the 
two sectors in the scaling limit can be defined separately leads to a 
consistent picture.) Applying $I^-$ on (\ref{A.2}) we get a state 
\bml
\bee
  \Big\vert\eta,\vartheta;\xi,\varphi;
  {\textstyle{5\over2}},
  {\textstyle{3\over2}}\Big\rangle=
  A_{1,0}\Big\vert\eta,\vartheta;
  {\textstyle{3\over2}},
  {\textstyle{1\over2}}\Big\rangle_r
  \Big\vert\xi,\varphi;{1},{1}\Big\rangle_l+
  A_{0,1}\Big\vert\eta,\vartheta;
  {\textstyle{3\over2}},
  {\textstyle{3\over2}}\Big\rangle_r
  \Big\vert\xi,\varphi;{1},{0}\Big\rangle_l\ .\label{A.3a}
\ee
According to our proposition the state with one $\theta$ is of the structure 
\bee
  \Big\vert\eta,\vartheta;\xi,\varphi;
  {\textstyle{3\over2}},
  {\textstyle{3\over2}}\Big\rangle=
  B_{1,0}\Big\vert\eta,\vartheta;
  {\textstyle{3\over2}},
  {\textstyle{1\over2}}\Big\rangle_r
  \Big\vert\xi,\varphi;{1},{1}\Big\rangle_l+
  B_{0,1}\Big\vert\eta,\vartheta;
  {\textstyle{3\over2}},
  {\textstyle{3\over2}}\Big\rangle_r
  \Big\vert\xi,\varphi;{1},{0}\Big\rangle_l\ .\label{A.3b}
\ee
\eml\bml
Acting on (\ref{A.3a}) and (\ref{A.3b}) by $I^-$ leads to 
\bee\ba{l}
  \Big\vert\eta,\vartheta;\xi,\varphi;{5\over2},{1\over2}\Big\rangle=\\
  A_{2,0}\Big\vert\eta,\vartheta;{3\over2},-{1\over2}\Big\rangle_r
  \Big\vert\xi,\varphi;{1},{1}\Big\rangle_l+
  A_{1,1}\Big\vert\eta,\vartheta;{3\over2},{1\over2}\Big\rangle_r
  \Big\vert\xi,\varphi;{1},{0}\Big\rangle_l+
  A_{0,2}\Big\vert\eta,\vartheta;{3\over2},{3\over2}\Big\rangle_r
  \Big\vert\xi,\varphi;{1},{-1}\Big\rangle_l
\ea\label{A.4a}
\ee
and
\bee\ba{l}
  \Big\vert\eta,\vartheta;\xi,\varphi;{3\over2},{1\over2}\Big\rangle=\\
  B_{2,0}\Big\vert\eta,\vartheta;{3\over2},-{1\over2}\Big\rangle_r
  \Big\vert\xi,\varphi;{1},{1}\Big\rangle_l+
  B_{1,1}\Big\vert\eta,\vartheta;{3\over2},{1\over2}\Big\rangle_r
  \Big\vert\xi,\varphi;{1},{0}\Big\rangle_l+
  B_{0,2}\Big\vert\eta,\vartheta;{3\over2},{3\over2}\Big\rangle_r
  \Big\vert\xi,\varphi;{1},{-1}\Big\rangle_l.
\ea\label{A.4b}
\ee
At the same time the state with two $\theta$s is of the form 
\bee\ba{l}
  \Big\vert\eta,\vartheta;\xi,\varphi;{1\over2},{1\over2}\Big\rangle=\\
  C_{2,0}\Big\vert\eta,\vartheta;{3\over2},-{1\over2}\Big\rangle_r
  \Big\vert\xi,\varphi;{1},{1}\Big\rangle_l+
  C_{1,1}\Big\vert\eta,\vartheta;{3\over2},{1\over2}\Big\rangle_r
  \Big\vert\xi,\varphi;{1},{0}\Big\rangle_l+
  C_{0,2}\Big\vert\eta,\vartheta;{3\over2},{3\over2}\Big\rangle_r
  \Big\vert\xi,\varphi;{1},{-1}\Big\rangle_l.
\ea\label{A.4c}
\ee
\eml
By further applications of $I^-$ we get 
\bml
\bee\ba{l}
  \Big\vert\eta,\vartheta;\xi,\varphi;{5\over2},-{1\over2}\Big\rangle=\\
  A_{3,0}\Big\vert\eta,\vartheta;{3\over2},-{3\over2}\Big\rangle_r
  \Big\vert\xi,\varphi;{1},{1}\Big\rangle_l+
  A_{2,1}\Big\vert\eta,\vartheta;{3\over2},-{1\over2}\Big\rangle_r
  \Big\vert\xi,\varphi;{1},{0}\Big\rangle_l+
  A_{1,2}\Big\vert\eta,\vartheta;{3\over2},{1\over2}\Big\rangle_r
  \Big\vert\xi,\varphi;{1},{-1}\Big\rangle_l,
\ea\label{A.5a}
\ee
\bee\ba{l}
  \Big\vert\eta,\vartheta;\xi,\varphi;{3\over2},-{1\over2}\Big\rangle=\\
  B_{3,0}\Big\vert\eta,\vartheta;{3\over2},-{3\over2}\Big\rangle_r
  \Big\vert\xi,\varphi;{1},{1}\Big\rangle_l+
  B_{2,1}\Big\vert\eta,\vartheta;{3\over2},-{1\over2}\Big\rangle_r
  \Big\vert\xi,\varphi;{1},{0}\Big\rangle_l+
  B_{1,2}\Big\vert\eta,\vartheta;{3\over2},{1\over2}\Big\rangle_r
  \Big\vert\xi,\varphi;{1},{-1}\Big\rangle_l,
\ea\label{A.5b}
\ee
\bee\ba{l}
  \Big\vert\eta,\vartheta;\xi,\varphi;{1\over2},-{1\over2}\Big\rangle=\\
  C_{3,0}\Big\vert\eta,\vartheta;{3\over2},-{3\over2}\Big\rangle_r
  \Big\vert\xi,\varphi;{1},{1}\Big\rangle_l+
  C_{2,1}\Big\vert\eta,\vartheta;{3\over2},-{1\over2}\Big\rangle_r
  \Big\vert\xi,\varphi;{1},{0}\Big\rangle_l+
  C_{1,2}\Big\vert\eta,\vartheta;{3\over2},{1\over2}\Big\rangle_r
  \Big\vert\xi,\varphi;{1},{-1}\Big\rangle_l;
\ea\label{A.5c}
\ee
\eml
\bml
\bea
&&\Big\vert\eta,\vartheta;\xi,\varphi;
  {\textstyle{5\over2}},
  -{\textstyle{3\over2}}\Big\rangle=
  A_{2,2}\Big\vert\eta,\vartheta;
  {\textstyle{3\over2}},
  -{\textstyle{1\over2}}\Big\rangle_r
  \Big\vert\xi,\varphi;{1},{-1}\Big\rangle_l+
  A_{3,1}\Big\vert\eta,\vartheta;
  {\textstyle{3\over2}},
  -{\textstyle{3\over2}}\Big\rangle_r
  \Big\vert\xi,\varphi;{1},{0}\Big\rangle_l\ ,\label{A.6a}\\
&&\Big\vert\eta,\vartheta;\xi,\varphi;
{\textstyle{3\over2}},
-{\textstyle{3\over2}}\Big\rangle=
  B_{2,2}\Big\vert\eta,\vartheta;
  {\textstyle{3\over2}},
  -{\textstyle{1\over2}}\Big\rangle_r
  \Big\vert\xi,\varphi;{1},{-1}\Big\rangle_l+
  B_{3,1}\Big\vert\eta,\vartheta;
  {\textstyle{3\over2}},
  -{\textstyle{3\over2}}\Big\rangle_r
  \Big\vert\xi,\varphi;{1},{0}\Big\rangle_l\ ;\label{A.6b}
\eea
\eml
and
\bee
  \Big\vert\eta,\vartheta;\xi,\varphi;
  {\textstyle{5\over2}},
  -{\textstyle{5\over2}}\Big\rangle=
  \Big\vert\eta,\vartheta;
  {\textstyle{3\over2}},
  -{\textstyle{3\over2}}\Big\rangle_r
  \Big\vert\xi,\varphi;{1},{-1}\Big\rangle_l\ .\label{A.7}
\ee
(The various coefficients should be proportional to the corresponding
Clebsh-Gordan coefficients.) All these states are of the 
same energy and momentum. In the space spanned by them 
we can choose for a basis the states 
\bee
  \Big\vert\eta,\vartheta;{3\over2},I^3_r\Big\rangle_r
  \Big\vert\xi,\varphi;{1},I^3_l\Big\rangle_l
  \quad I^3_r={3\over2},{1\over2},-{1\over2},-{3\over2};\quad I^3_l=
  1,0,-1.
  \label{A.8}
\ee
These basis vectors  can be expressed in principle through 
eqs.~(\ref{A.2})-(\ref{A.8}). It is obvious, that this procedure can be performed 
for any 
$(\eta,\vartheta;\xi,\varphi)$ 
solution of Eqs.\ (\ref{3.8a},\ref{3.8b}) and (\ref{3.9a},\ref{3.9b}), 
thus the space spanned by the 
scaling limits of the solutions for the BA equations factorises into the
direct product of the spaces spanned by the isospin eigenstates of 
the `r' and `l' sectors. 

\section{}\label{sec:delta}

Deriving the HLBA equations one neglects corrections exponentially
small in the chainlength (terms of the type 
$\delta\propto\exp\{-\alpha(u)N\}$) at two points:
1) when describing the bound pairs of (\ref{2.3}) by a single parameter 
(the $\Lambda$), and 2) when representing the complex $\Lambda$s by the set
$\Theta$. Although in the scaling limit $N\to\infty$, $\alpha(u)$ may
tend to zero, thus one has to check if these corrections remain 
negligible indeed in the scaling limit. 
This is what we do in this Appendix.

1)\ First we discuss the correction terms to the size (in wavenumber space)
of the bound pairs in the scaling limit. The relation (\ref{2.3}) completed 
with the correction terms reads
\bee
 \sin k^{\pm}=\Lambda\pm iu+\delta^{\pm}\,.\label{2.3app}
\ee
In (\ref{2.3app}) $\delta$ can be estimated 
in the following way. Substituting
(\ref{2.3app}) into (\ref{2.1a}) we have (up to leading order in $\delta$) 
\bee
\ln{\vert\delta\vert\over2u}=
-{\rm Im}\left\lbrace Nk^+-\sum_{\Lambda^{\prime}(\not=\Lambda)}2\tan^{-1}
{\Lambda+iu-\Lambda^{\prime}\over u}\right\rbrace\,.
\ee
The sum over the $\Lambda^{\prime}$ (in leading order in $N$) can be 
calculated by means of the ground-state density of $\Lambda$:
\bee\label{lndelta}
\ln{\vert\delta\vert\over2u}=
-N{\rm Im}\left\lbrace k^+-\int\limits_{-\infty}^{\infty}2\tan^{-1}
{\Lambda+iu-\Lambda^{\prime}\over u}\,\sigma_0(\Lambda^{\prime})
d\Lambda^{\prime}\right\rbrace\,,
\ee
where (according to \cite{Wo2}):
\bee\label{szigma0}
\sigma_0(\Lambda)={1\over2\pi}\int\limits_0^{\infty}
{J_0(\omega)\cos(\omega\Lambda)\over\cosh(\omega u)}d\omega\,,
\ee
with $J_0$ being the zeroth order Bessel function. 
As for $u\ll1$ the main contribution of the integral comes from the 
region $\vert\Lambda^{\prime}\vert<1$,
evaluating the r.h.s.\ of (\ref{lndelta})
the regions $\vert\Lambda\vert>1$ and $\vert\Lambda\vert<1$ have to
be treated differently (for the sake of simplicity we consider here real 
$\Lambda$s only; a complex $\Lambda$ can be treated in a similar way).
In the first case for $u\ll1$ the integral is real, 
(${\rm sgn}(\Lambda)\pi/2$) giving no contribution to the imaginary part, 
and then the
task is simply to calculate the ${\rm Im}\,k^+$ 
\bee
{\rm Im}\,k^+=\cosh^{-1}\left\lbrace{1\over2}\left(\sqrt{u^2+(\Lambda+1)^2}
+\sqrt{u^2+(\Lambda-1)^2}\right)\right\rbrace\,.
\ee
For the second case the integral can not be neglected, and we have to 
calculate 
\bea
&&{\displaystyle \cosh^{-1}\left\lbrace{1\over2}\left(\sqrt{u^2+(\Lambda+1)^2}
+\sqrt{u^2+(\Lambda-1)^2}\right)\right\rbrace}-\nonumber\\
&&\quad\quad\quad\quad\quad\quad\quad\quad
-{\displaystyle {1\over8\pi}\int\limits_{-\infty}^{\infty}
\int\limits_{-\infty}^{\infty}
\ln{(\Lambda-\Lambda^{\prime})^2\over
4u^2+(\Lambda-\Lambda^{\prime})^2}\,
{J_0(\omega)\exp(i\omega\Lambda^{\prime})\over\cosh(\omega u)}}\,.
\eea
This expression can be evaluated (approximately) for small $u$.
Finally we obtain
\bee
\ln{\vert\delta\vert\over2u}=\left\lbrace\ba{ll}
-{\displaystyle N\cosh^{-1}\vert\Lambda\vert}&
\mbox{$\quad\quad(\vert\Lambda\vert>1)$}\,,\\
-{\displaystyle N{4\over\pi}K_0\left({\pi\over2u}\right)
\cosh{\pi\Lambda\over2u}}&
\mbox{$\quad\quad(\vert\Lambda\vert<1)$}\,,
\ea\right.
\ee
with $K_0$ being the modified Bessel function. This shows, that in the 
scaling limit (\ref{1.3}) $\delta\to0$ everywhere except in a small
vicinity of $\Lambda=0$. There it behaves as 
\bee\label{deltak2}
\vert\delta\vert=2u
\exp\left\lbrace-Lm_0\cosh{\pi\Lambda\over2u}\right\rbrace\,.
\ee
Consequently the bound pairs are of the form (\ref{2.3}) indeed up to
exponentially small corrections. Moreover, at the ends of the 
$\Lambda$ distribution, where the holes corresponding to the massless 
particles are situated, (\ref{2.3}) becomes exact in the scaling limit.

It is apparent that in the scaling limit the $\delta$s 
in (\ref{deltak2}) can be larger than $1/N$. 
Nevertheless their effect 
is negligible compared to the {\em leading} finite size corrections,
which are actually of the order of $1/L$ 
(see Sec.~\ref{sec:conform} and Appendix \ref{sec:vegesm}), 
as the coefficients of the $1/N$
terms are of the order of $N/L$. 

2)\ Now we discuss the neglected corrections to the complex $\Lambda$s. The 
complex $\Lambda$s are represented by the set $\Theta$ in the following way:
a $\Theta$ with an imaginary part $\vert{\rm Im}\Theta\vert>u$ represents a
wide root:
\bml
\bee
\Theta\longrightarrow \Lambda=\Theta+iu\,{\rm sgn}({\rm Im}\Theta)\quad
(\vert{\rm Im}\Theta\vert>u)\,,
\ee
and a $\Theta$ with an imaginary part $\vert{\rm Im}\Theta\vert<u$ represents
a close {\em pair}\/:
\bee\label{szukpar}
\Theta\longrightarrow \Lambda^{\pm}=\Theta\pm iu\pm\tau\quad
(\vert{\rm Im}\Theta\vert<u)\,,
\ee\eml
with $\tau$ expected to be exponentially small.
Its magnitude can be estimated on the basis, that 
$\Lambda^{\pm}$ must satisfy Eq.\ (\ref{2.5}).
Substituting $\Lambda^+$ into (\ref{2.5}) we
have
\bea\label{kozeli}
2\pi J&=&N\left(\sin^{-1}\!(\Lambda^+-iu)+
  \sin^{-1}\!(\Lambda^++iu)\right)
  -\sum_k2\tan^{-1}{\Lambda^+-\sin k\over u}\,-\nonumber\\
  &-&\sum_{\Lambda^{\prime}(\not=\Lambda^-)}2\tan^{-1}{\Lambda^+-
  \Lambda^{\prime}\over 2u}-2\tan^{-1}{\Lambda^+-
  \Lambda^-\over 2u}\ .
\eea
Neglecting the effect of the $k$s (as being not macroscopic) and
estimating the sum over the $\Lambda^{\prime}$ by means of the ground-state
$\Lambda$ distribution $\sigma_0$, (\ref{kozeli}) yields to 
leading order in $\tau$
\bee
  -\ln{\vert\tau\vert\over2u}=N{\rm Im}\left\lbrace
  \left(\sin^{-1}\!(\Lambda^+-iu)+\sin^{-1}\!(\Lambda^++iu)\right)
  -\int_{-\infty}^{\infty}2\tan^{-1}{\Lambda^+-
  \Lambda^{\prime}\over 2u}\sigma_0(\Lambda^{\prime})\right\rbrace\ .
\ee
The r.h.s.\ for small $u$ can be evaluated in a closed form
if the $\vert{\rm Re}\Lambda^+\vert>1$. The result for
${\rm Re}\Lambda^+>1$ reads:
\bee
N2\sin{\pi{\rm Im}\Lambda^+\over2u}I_0\left({\pi\over2u}\right)
\exp\left\lbrace{-{\pi{\rm Re}\Lambda^+\over2u}}\right\rbrace\,,
\ee
which in the scaling limit yields
\bee
{\vert\tau\vert\over2u}=\exp\left\lbrace-
Lm_0\cos{\rm Im}\vartheta\,e^{{\rm Re}\vartheta}\right\rbrace\,,
\ee
with $\vartheta$ (in agreement with (\ref{3.6}))
\bee
\vartheta=-{\pi(\Lambda^+-iu-2)\over2u}\,.
\ee
The analogous expression for the ${\rm Re}\Lambda^+<-1$ (which
are connected to the `l' particles)
\bee
{\vert\tau\vert\over2u}=\exp\left\lbrace-
Lm_0\cos{\rm Im}\varphi\,e^{-{\rm Re}\varphi}\right\rbrace\,,
\ee
with
\bee
\varphi=-{\pi(\Lambda^+-iu+2)\over2u}\,.
\ee

The conclusion of the above is, that the corrections to the size of the
close pairs are negligible indeed as long as  
${\rm Re}\,\vartheta\gg-\ln(Lm_0)$
and ${\rm Re}\,\varphi\ll\ln(Lm_0)$. ${\rm Re}\,\vartheta\simeq-\ln(Lm_0)$
and ${\rm Re}\,\varphi\simeq\ln(Lm_0)$ can occur only in the smallest 
energy states
(when the $\eta\simeq-\ln(Lm_0)$
and $\xi\simeq(Lm_0)$), i.e.~one has to treat only the smallest energy states
in a more subtle way, all the others are correctly described by 
Eqs.~(\ref{3.8}) and (\ref{3.9}).  

\section{}\label{sec:vegesm}

In this Appendix we would like to discuss the finite size corrections 
in more details. 
The procedure given in Sec.~\ref{sec:conform} 
is based on the fact, that the energy-momentum dispersion relation 
of the massless
(dressed) particles is linear, and that the dressed particles are 
described by BA type equations. Here we describe the better known method
\cite{Wo4,HaQuBa}, 
which does not use these properties directly.

Our starting point is given by the Eqs.~(\ref{2.4})-(\ref{2.5}) and the 
dispersion relation (\ref{2.6}).
For the sake of simplicity we consider states with no massive particles.
The $\Lambda$ distribution is supposed to be the simplest, i.e.~the 
$J_{\eta}$ parameters form an equidistant series between 
$J^+\leq(N-n(\Lambda)-1)/2$ and $J^-\geq-(N-n(\Lambda)-1)/2$ 
(with $J^+-J^-=n(\Lambda)-1$), i.e.:
\bee
J^-=J_1,\quad J_{\eta}+1=J_{\eta+1},\quad J_{n(\Lambda)}=J^+\,.
\ee
After introducing the function
\bee
z(\Lambda)={1\over2\pi}\left\lbrace\left(\sin^{-1}\!(\Lambda-iu)+
  \sin^{-1}\!(\Lambda+iu)\right)-{1\over N}\sum_{\nu=1}^{n(\Lambda)}
  2\tan^{-1}{\Lambda-
  \Lambda_{\nu}\over 2u}\right\rbrace\,,
\ee
through the equation
\bee
J_{\eta}/N=z(\Lambda_{\eta})\,
\ee
one can interpret the $\Lambda_{\eta}$ as a function of $J_{\eta}$.
The density of the roots is given by
\bee
\sigma(\Lambda)={d\,z(\Lambda)\over d\,\Lambda}\,.
\ee
Defining 
\bee
S(\Lambda)={1\over N}\sum_{\eta}\delta(\Lambda-\Lambda_{\eta})-
\sigma(\Lambda)\,,
\ee
we can write
\bea
\sigma(\Lambda)=&&{\displaystyle {1\over2\pi}\Biggl\lbrace
\left(\left(1-(\Lambda-iu)^2\right)^{-1/2}+
  \left(1-(\Lambda+iu)^2\right)^{-1/2}\right)\Biggr.}\nonumber\\
 &&\quad\quad\quad{\displaystyle\Biggl.-\int_{-\infty}^{+\infty}
 {4u\over4u^2+(\Lambda-\Lambda')^2}
  (\sigma(\Lambda')+S(\Lambda'))\Biggr\rbrace}\,.
\eea
This equation, after some straightforward algebra leads to
\bee\label{suruseg}
\sigma(\Lambda)=\sigma_{0}(\Lambda)-\int\limits_{-\infty}^{+\infty}
\bar K(\Lambda-\Lambda')S(\Lambda')\,,
\ee
with $\sigma_0$ given by (\ref{szigma0}) and
\bee
\bar K(\Lambda)={1\over2\pi}\int\limits_{-\infty}^{+\infty}
{e^{i\omega\Lambda}\over1+e^{2\vert\omega\vert u}}\,.
\ee
The energy
\bee
E=-Nu
  -\sum_{\eta}2t\left(\sqrt{1-(\Lambda_{\eta}-iu)^2}
  +\sqrt{1-(\Lambda_{\eta}+iu)^2}-2u\right)\,,
\ee
after some manipulation turns out to be 
\bee
E=N\varepsilon_0-\int\limits_{-\infty}^{+\infty}
\varepsilon_c(\Lambda)S(\Lambda)\,,
\ee
with $\varepsilon_0$ and $\varepsilon_c$ given by (\ref{2.10}).
To calculate integrals of $S(\Lambda)$ the Euler-Maclaurin type formula
\bee\label{EuMac}
{1\over2N}f\left(J_1\over N\right)+
{1\over N}\sum_{\eta=2}^{n-1}f\left(J_{\eta}\over N\right)+
{1\over2N}f\left(J_n\over N\right)=
\int\limits_{J_1/N}^{J_n/N}f\left(x\right)+
{1\over12N^2}\left(f'\left(J_n\over N\right)-f'\left(J_1\over N\right)\right)
\ee 
($f'$ being the derivative of $f$) can be used. 
(We note here, that (\ref{EuMac}) applied in this and related problems 
(like the Heisenberg chain) takes into account the leading terms of a
nonconvergent (asymptotic) expansion. Nevertheless the result is reliable,
as it is well indicated by the numerical calculations for very long chains 
\cite{Wo4}.) This leads to
\bee
E-N\varepsilon_0=
N\varepsilon^++N\varepsilon^-\,,
\ee
where
\bee
\varepsilon^+=\int\limits_{\Lambda^+}^{+\infty}
\varepsilon(\Lambda)\sigma(\Lambda)-{1\over2N}\varepsilon(\Lambda^+)
-{\varepsilon'(\Lambda^+)\over12N^2\sigma(\Lambda^+)}\,,
\ee
with $\Lambda^+$ being the largest root:
\bee
{J^+\over N}=z(\Lambda^+)\,,\quad\left({\rm i.e.:}\quad 
{N-n(\Lambda)\over N}-{J^+\over N}=\int\limits_{\Lambda^+}^{+\infty}
\sigma(\Lambda^{\prime})\,\right),
\ee
and analogous expressions give $\varepsilon^-$ and $\Lambda^-$.
It is apparent, that to calculate $\varepsilon^+$, it is enough to know 
$\sigma(\Lambda)$ for $\Lambda^+\leq\Lambda<+\infty$. Applying (\ref{EuMac})
in (\ref{suruseg}) and neglecting the next to leading terms one arrives
at the equation
\bee\label{WH}
\sigma(\Lambda)=\sigma_{0}(\Lambda)+\int\limits_{\Lambda^+}^{+\infty}
\bar K(\Lambda-\Lambda')\sigma(\Lambda')-{1\over2N}{\bar K}(\Lambda-\Lambda^+)
+{{\bar K}^{\prime}(\Lambda-\Lambda^+)\over12N^2\sigma(\Lambda^+)}
\ee
wich determines $\sigma$ for $\Lambda>\Lambda^+$ with sufficient
accuracy. (\ref{WH}) is a Wiener-Hopf type equation, and can 
be solved in closed form 
\cite{Wo4,HaQuBa} leading to
\bee
\varepsilon^+=8\pi t{I_1\left({\pi\over2u}\right)\over 
I_0\left({\pi\over2u}\right)}\left\lbrace{1\over8}{(n^+)^2\over N^2}-
{1\over48N^2}\right\rbrace\,,\quad n^+={N-n(\Lambda)-1\over2}-J^+\,.
\ee
The analogous expression for $\varepsilon^-$ is
\bee
\varepsilon^-=8\pi t{I_1\left({\pi\over2u}\right)\over 
I_0\left({\pi\over2u}\right)}\left\lbrace{1\over8}{(n^-)^2\over N^2}-
{1\over48N^2}\right\rbrace\,,\quad n^-={N-n(\Lambda)-1\over2}+J^-\,.
\ee
The momentum $P=\sum J_{\eta}/N$ through an elementary calculation is found
to be
\bee
P={\pi\over2}(n^+-n^-)+{2\pi\over N}\left\lbrace{(n^-)^2\over4}-{(n^+)^2\over4}
\right\rbrace\,.
\ee
Completing the scaling limit, and renaming $n^{\pm}$ one arrives at
\bee
E-N\varepsilon_0=-{\pi\over6L}+{2\pi\over L}
\left\lbrace{(n_r)^2\over4}+{(n_l)^2\over4}
\right\rbrace\,,\quad
P={2\pi\over L}\left\lbrace{(n_r)^2\over4}-{(n_l)^2\over4}
\right\rbrace\,.
\ee
This is the same as obtained in Sec.~\ref{sec:conform}.

\section{}\label{sec:fren}

Here we outline the main results of the perturbative computation
of the free energy in the CGN model \cite{FoNi}.
First one introduces a chemical potential, $h$, coupled to a Noether charge
of the U($2$) symmetry, $Q$,
and considers the ground state of the system with the Hamiltonian
\bee \label{hamil}
 H=H_{0} - hQ\,.
\ee
The charge matrix $Q$ has the 
eigenvalues $Q_a$ on the bare fields:
$Q\psi_a=Q_a\psi_a$.
By decomposing this diagonal matrix into abelian (SU(2) singlet)
and non--abelian (traceless) parts one can write:
\bee\label{qs}
Q_{ab}=q\delta_{ab}+Q\tau^3_{ab} \,\quad \tau^i={1\over2}\sigma^i\,,
\ee
with $\sigma^i$ ($i=$1,2,3) being the usual Pauli matrices.
As mentioned above, in the CGN model the massive particles couple only to the
non--abelian part of the charge, $Q$, while the massless sector couples
only to the abelian component, $q$.
Consequently, the free energy (the ground state energy of
(\ref{hamil})) is expected to have the form:
\bee \label{free0}
\df \equiv f(h)-f(0) = \bar{f}(h,q,g')+
    \hat{f}(h,Q,g(h))\,,
\ee
where $g'=g_v+g_s/2$ is the coupling of the abelian sector.
Perturbation theory is applicable for weak coupling, 
which implies high densities (asymptotic freedom) as the running
coupling, $g(h)$, behaves as $g(h)\propto \lb \ln (h/\Lambda) \rb^{-1}$,
where  
$\Lambda$ is a parameter
(of dimension mass) which is a renormalization group invariant combination
of the ultraviolet cutoff and the bare coupling.
The free energy of the massive sector, $\hat{f}$,
has in fact the
form $\hat{f}=h^2 \Phi (h/\Lambda,Q)$ and by 
expressing the running
coupling, $g(h)$, in terms of the $\Lambda$-parameter one obtains
an asymptotic series in $[\ln( h/\Lambda)]^{-1}$ for
$\Phi(h/\Lambda,Q)$.
As the Abelian coupling $g'$ has a vanishing beta function
$\bar{f}(h,q,g')$ is simply given as $h^2\phi(q,g')$.
The calculation of the free energy in the scaling limit of the Hubbard model
based on the HLBA eqs.\  yields a result of the form
$\delta f(h)=h^2\Psi (h/m_0)+h^2{\rm const}$,
and when for a large density $\Psi(h/m_0)$ is expanded
into an asymptotic series in $1/[\ln(h/m_0)]$,
the results for 
$\Phi$ and $\Psi$ should agree as one calculates the same quantity
provided the two models are indeed equivalent.
One needs however to relate the two mass scales $m_0$ and $\Lambda$
in order to achieve agreement.
  
The abelian part of the free energy, $\bar{f}$, receives contribution only
from bubble diagrams whose sum yields:
\bee \label{fbarp}
\bar{f}=-{h^2q^2 \over \pi} {1 \over 1+ 2 g' / \pi} \,.
\ee
This can be immediately seen to agree with the result 
(\ref{fre}) found in Section V
for the massless sector by identifying the chemical potentials
provided that we set $g'=0$.  

The non--abelian part of the perturbative free energy
had to be calculated up to three loop order with the result:
\bee \label{fhatp1}
\hat{f} = -\frac{h^2}{2\pi}\frac{Q^2}{2} 
\left\lb 1 - \frac{g(h)}{\pi}
+ C \frac{g^2(h)}{\pi^2} +O(g^6) \right\rb,
\ee
where $g(h)$ is the renormalized running coupling 
in the $\MSb$ scheme and the constant $C$ is given as
\bee \label{c}
C = 2\ln2-\h+2\ln Q\,.
\ee
Introducing now the renormalization group invariant
$\Lambda_{\MSb}$--parameter
corresponding to the running coupling:
\bee \label{gh}
g(h) = {1 \over \beta_0 \lnhb} -
{\beta_1 \ln\lnhb \over \beta_0^3(\lnhb)^2}
+O \left({\ln\lnhb \over (\lnhb)^3}\right)\,,
\ee 
where $\beta_0$ and $\beta_1$ are the
universal 1-- and 2--loop beta--function
coefficients.
The beta-function of the SU(2) CGN model up to two loops is given as
\cite{De}: 
\bee \label{rgeq}
h {\partial \over \partial h} g(h) =
 - \beta_0 g^{2}(h) - \beta_1 g^{3}(h) - \ldots \,,
\ee
\bee \label{cgnbf}
\beta_{0} =  \frac{2}{\pi},
\qquad \beta_{1} =  -\frac{2}{\pi^2} \,.
\ee
In order to facilitate the comparision of the perturbative
free energy with the nonperturbative result derived from the HLBA
eqs.\ of the Hubbard model we rewrite 
the perturtbative free energy (\ref{gh}) using 
$\lnhb=t+\ln R_{\MSb}$, where $R_{\MSb}=m_0/\Lambda_{\MSb}$:
\bee \label{fp2}
\hat f(h) = -\frac{h^2}{2\pi}\cdot { Q^2\over2}
\left\{ 1 - \frac{1}{2 t}
-\frac{\ln t}{4t^2}
+ \frac{1}{2 t^2}\tilde C+O\left({\ln t\over t^3}\right)
\right\}\,, \ee
where
\bee \label{tc}
\tilde C
=\frac{1}{2}C+\ln R_{\MSb}
\ee

\section{}\label{sec:nlim}

We sketch here briefly the `naive' continuum limit of the half filled
Hubbard chain.
A more detailed description of the procedure is given in Ref.\ \cite{WoEcTr}.

Let us redefine the chain so, that four lattice sites form one elementary
cell and define the operators
\bee
\phi_{\nu,\sigma}(n)={1\over4\sqrt{a}}\sum_{j=1}^4(-i)^{\nu j}
c_{4(n-1)+j,\sigma}\quad \nu=1,2,3,4,\quad\sigma=\uparrow,\downarrow.
\ee
In terms of these the original Fermion operators are
\bee\label{cek}
c_{4(n-1)+l,\sigma}=\sqrt{a}\sum_{\nu=1}^4(i)^{\nu l}\phi_{\nu,\sigma}(n)
\quad l=1,2,3,4.
\ee
The length of the chain is 
\bee
L=Na=4aN'\ (N'=int.)
\ee 
and the continuum limit is defined as
$$a\to0,\quad N'\to\infty,\quad L=fixed$$
with the continuous variable 
\bee
x=4a(n-1/2),\quad dx=4a.
\ee
In this limit
\bee\label{sumint}
\sum_n^{N'}\to\int\limits_0^L{dx\over4a}
\ee
\bee
\delta_{n,n'}\to4a\delta(x-x').
\ee
If we introduce
\bee\label{pszik}
\phi_{1,\sigma}(n)=\psi_{1,\sigma}(x),
\quad\phi_{3,\sigma}(n)=\psi_{2,\sigma}(x),
\ee
\bee
\left\{\psi_{\alpha,\sigma}(x),\psi_{\beta,\sigma'}^{+}(x')\right\}=
\delta(x-x')\delta_{\alpha,\beta}\delta_{\sigma,\sigma'}.
\ee
Applying (\ref{cek}), (\ref{pszik}) and (\ref{sumint}) to the Hubbard 
Hamiltonian (\ref{1.1}) and neglecting $\phi_2$ and $\phi_4$ (this can be 
done as these fields become infinitely massive in the continum limit)
one obtains a Hamiltonian which is not even SU(2) symmetric.
Observing, however, that the interaction part of the Hamiltonian 
(\ref{1.1}) can also be written as
\bee\label{hprime}
\hat H_{int}=H'-U{\hat N}+{1\over4}UN
\quad\hbox{where}\quad H'={U\over2}\sum_{j=1}^N 
(n_{j\uparrow}+n_{j\downarrow})^2\,
\quad\hbox{and}\quad \hat N=\sum_{j=1}^N 
(n_{j\uparrow}+n_{j\downarrow})\,,
\ee
one obtains in the naive continuum limit the following Hamiltonian density:
\bee
{\cal H}(x)=(2at)\left(-\sum_{\sigma=1}^2\psi_{\sigma}^+
\gamma_5\pa_x\psi_{\sigma}+
u\left[(\sum_\sigma\psi_{\sigma}^+\gamma_0\psi_{\sigma})^2
+(\sum_\sigma\psi_{\sigma}^+\psi_{\sigma})^2\right]\right)
\ee
where
\bee
\psi_{\sigma}=\pmatrix{\psi_{1,\sigma}\cr
\psi_{2,\sigma}\cr}\,,\qquad
\gamma_5=\pmatrix{i&0\cr            
                  0&-i\cr}\,,\qquad 
 \gamma_0=\pmatrix{0&1\cr            
                  1&0\cr}\,,                   
\ee                   
and $u=U/4t$. (Note that this differs from the definition used in the 
bulk of the paper as there $u=|U|/4t$.)

\section{}
\label{sec:coupl}
In this last Appendix we would like to present a one loop
renormalization group analysis of the naive continuum limit of the half 
filled Hubbard model (\ref{naivlim}).

Let us consider the following non Lorentz invariant 
theory with four fermion interactions:
\bee \label{generalth}
{\cal L} =  i\bar{\psi_a} \delsl \psi_a
   -{g_0\over2}(\bar{\psi_a} \psi_a)^2
   -{g_{10}\over2}(\bar{\psi_a} \gamma_{0} \psi_a)^2
   -{g_{11}\over2}(\bar{\psi_a} \gamma_{1} \psi_a)^2
   -{g_{2}\over2}(\bar{\psi_a} \gamma_{5} \psi_a)^2  \,,
\ee
The theory given by (\ref{generalth}) is clearly renormalizable at least up
to one loop  and contains the theory obtained in the naive continuum limit
(\ref{naivlim}).
The CGN model now corresponds to the choice of
$g_0=g_{2}=2g_{10}=2g_{11}$, while the nonchiral
GN model is obtained as $g_{2}=g_{10}=g_{11}=0$.
The theory found in the naive continuum limit (\ref{naivlim}) corresponds
to 
\bee\label{inicond}
g_0=g_{10}=2u$,\quad $g_2=g_{11}=0\,.
\ee
Our strategy is to 
compute the one loop $\beta$-functions in the
space of the four couplings $\{g_0\,,g_{10}\,,g_{11}\,,g_2\}$
and then investigate if the classically neither Lorentz nor chiral invariant
theory (\ref{naivlim}) could somehow be connected to the SU(2) CGN
model in the one loop approximation. 
The calculation of the one loop $\beta$-functions in perturbation theory
is by now standard and therefore
we just state the result of the calculation without going into the details.
\bee\label{4bf}\eqalign{
\beta_0&=-{1\over\pi}\left[(n-1)g_0^2+g_0g_2-(g_{10}+g_{11})(g_0-g_2)\right]\,,\cr
\beta_2&=-{1\over\pi}\left[(n-1)g_2^2+g_0g_2+(g_{10}+g_{11})(g_0-g_2)\right]\,,\cr
\beta_{10}&=-{1\over\pi}g_0g_2\,,\cr
\beta_{11}&=-{1\over\pi}g_0g_2 \,,}
\ee
where $\beta_i=\mu{dg_i/d\mu}$.
The calculation has been performed for the SU(n)
symmetric case in the dimensional regularisation
scheme.  One avoids the problem of evanescent operators
since in the computation of the
$\beta$-functions up to one loop order, only the divergent pieces are needed
hence
one can perform the Dirac algebra
in two dimensions. For the Lorentz invariant case $g_{10}=g_{11}$
Eqs.~(\ref{4bf}) reproduce the known result of Ref.~\cite{MiWe}. 
For the present case of interest $n=2$ and then one can easily solve 
Eqs.~(\ref{4bf}) for arbitrary initial conditions.
Starting with the initial conditions (\ref{inicond}) at $t=0$ the evolution
of the couplings is given as:
\bee\label{evol}\eqalign{
g_0&=u\left({1\over1+2ut/\pi}+{1\over1-2ut/\pi}\right)\cr
g_2&=u\left({1\over1+2ut/\pi}-{1\over1-2ut/\pi}\right)\cr
g_{10}&={u\over2}\left({1\over1+2ut/\pi}+{1\over1-2ut/\pi}\right)+u\cr
g_{11}&={u\over2}\left({1\over1+2ut/\pi}+{1\over1-2ut/\pi}\right)-u}
\ee
Independently on the sign of the coupling of the Hubbard model, $u$,
the couplings $\{g_0,g_{10},g_{11},g_2\}$
will inevitably become large for $t\rightarrow\pm\infty$ (both in the
ultraviolet (UV) and in the infrared (IR)) and perturbation theory then breaks
down. In other words one hits a Landau pole both in the UV and in the
IF region which appears to be rather peculiar.
For $u>0$ in the $t\rightarrow\infty$ limit the couplings will
approximatively satisfy
\bee\label{pos}
 g_0\approx u\approx -g_2\approx 2g_{10}\approx2g_{11}\,,
 \ee
while for $u<0$
\bee\label{neg}
 g_0\approx u\approx g_2\approx 2g_{10}\approx2g_{11}\,.
\ee
In the IF limit the situation is reversed in that then 
one  gets close to the CGN trajectory for $u>0$.
Unfortunately the one loop beta-function of the Lorentz-invariance
violating interaction ($g_{10}-g_{11}$) is zero, so 
in the one loop approximation no definite
statement about the restoration of Lorentz invariance can be made. 
Clearly it is tempting to conclude from  Eqs.\ (\ref{pos},\ref{neg}) 
that Lorentz invariance
will be restored and that the IF limit of theory (\ref{naivlim}) 
for $u>0$ resp.\ its UV limit for $u<0$ will eventually be the
CGN model.
As the one loop approximation is applicable, however, only for
$t\ll1/\vert u\vert$,
the above can only be considered as an indication about the limit
of (\ref{naivlim}). Nevertheless it
is quite interesting that this might be an explicit example
of the restoration of Lorentz invariance in the quantum theory.


\newpage
\begin{references}

%
\bibitem{Sha}
B.S.~Shastry; J.Stat.Phys.{\bf50} (1988) 57
%
\bibitem{LiWu}
E.H.~Lieb, F.Y.~Wu; Phys.Rev.Lett.{\bf20} (1968) 1445
%
\bibitem{Exactly}for a good collection of reprints and comments see 
V.E.~Korepin, F.H.L.~E\ss\-ler; 
Exactly solvable models of strongly correlated electrons, World Scientific,
Singapore, 1994.
%
\bibitem{Ya} 
C.N.~Yang; Phys.Rev.Lett.{\bf 63} (1989) 2144,\\ 
C.N.~Yang, S. Zhang; Mod.Phys.Lett.~{\bf B4} (1990) 759
%
\bibitem{Per}
M.~Pernici; Europhys.Lett. {\bf 12} (1990) 75
%
\bibitem{EssKo} 
F.H.L.~Essler, V.E.~Korepin; Nucl.Phys.B{\bf 426} (1994) [FS] 505
%
\bibitem{Wo1} 
F.~Woynarovich; J.Phys.C{\bf16} (1983) 5293
%
\bibitem{Wo2} 
F.~Woynarovich; J.Phys.C{\bf16} (1983) 6593
%
\bibitem{Ov} 
A.A.~Ovchinnikov; Sov.Phys.JEPT {\bf 30} (1970) 1160
%
\bibitem{Me} 
E.~Melzer; Nucl.Phys.B443[FS] (1995) 553
%
\bibitem{Fi} 
V.M.~Filev; Teor.i Mat.Fiz,{\bf 33} (1977) 119
%
\bibitem{GrNe} 
D.~Gross and A.~Neveu; Phys.Rev.D{\bf 10} (1974) 3235
%
\bibitem{MiWe}
P. K. Mitter and P. H. Weisz,
Phys. Rev. D8 (1973) 4410.
%
\bibitem{BKKW}
B. Berg, M. Karowski, V. Kurak and P. Weisz,
Nucl. Phys. B134 (1978) 125.
%
\bibitem{ABW}
B. Berg and P. Weisz,
Nucl. Phys. B146 (1979) 205; \\
E. Abdalla, B. Berg and P. Weisz,
Nucl. Phys. B157 (1979) 387.
%
\bibitem{KoKuSw}
R. K\"oberle, V. Kurak and J. A. Swieca,
Phys. Rev. D20 (1979) 897;\\ 
Erratum -- {\it ibid} D20 (1979) 2638.
%
\bibitem{Wit}
E. Witten; Nucl. Phys. B145 (1978) 110.
%
\bibitem{AnLo1} N.~Andrei, J.H. Lowenstein; 
Phys.Rev.Lett.{\bf43} (1979) 1698,
%
\bibitem{Bel}
A. A. Belavin; Phys. Lett. 87B (1979) 117.
%
\bibitem{AnLo2} N.~Andrei, J.H. Lowenstein;
Phys.Lett. {\bf 91B} (1980) 401 
%
\bibitem{Zam2} 
A.~B.~Zamolodchikov and Al.~B.~Zamolodchikov; Nucl.Phys.B{\bf 379} (1992) 
 602
\bibitem{Aff}
I.~Affleck; talk given at the NATO Advanced Study Institute on
Physics, geometry and topology, Banff (August 1989) 
%
\bibitem{FoNi}
P.~Forgacs, S.~Naik, F.~Niedermayer, Phys.~Lett. 283B (1992) 282.
%
\bibitem{EssKoSchou} 
F.H.L.~Essler, V.E.~Korepin, K.L~Schoutens; Nucl.Phys.B{\bf 384} (1992) 431
%
\bibitem{Wo3} 
F.~Woynarovich; J.Phys.C{\bf15} (1982) 85
%
\bibitem{KluSch} 
A.~Kl\"umper, A. Schadschneider, J. Zittartz; Z.Phys.B{\bf78} (1990) 99
%

\bibitem{Kor}
V.E.~Korepin; Theor.~Math.~Phys.~76 (1980) 165
%
\bibitem{DeLo1} 
C.~Destri, J.H.~Lowenstein; Nucl.Phys.B205[FS5] (1982) 369
%
\bibitem{DeLo2} 
C.~Destri, J.H.~Lowenstein; Nucl.Phys.B200[FS4] (1982) 71
%
\bibitem{Wo4} 
F.~Woynarovich, H-P.~Eckle; J.Phys.A{\bf20} (1987) L443
%
\bibitem{Car}
Phase Transitions and Critical Phenomena, ed.\ C.Domb, J.L.Lebowitz;
Academic, New York, 1987.
%
\bibitem{KlPe}
A.~Kl\"umper, P.~A.~Pearce; Phys.Rev.Lett.~{\bf 66} (1991) 974
%
\bibitem{PeKl}
P.~A.~Pearce, A.~Kl\"umper; J.Stat.Phys. {\bf 64} (1991) 13
%
\bibitem{FoNiWe}
P.~Forg\'acs, F.~Niedermayer and P.~Weisz,
Nucl. Phys. B367 (1991) 123; 
%
\bibitem{So}
J.~S\'olyom; Advances in Physics {\bf 28} (1979) p.201
%
\bibitem{MoScha}
E. Moreno and F. A. Schaposnik,
Int. J. of Mod. Phys, A4 (1989) 2827.
%
\bibitem{De}
C. Destri, Phys.~Lett.~210B (1988) 173;
Erratum -- {\it ibid} 213B (1988) 565.
%
\bibitem{HaQuBa}
C.~J.~Hamer, G.~R.~W.~Quispel, M.~T.~Batchelor; J.Phys.A {\bf20} (1987) 5677 
%
\bibitem{WoEcTr}
F.~Woynarovich, H-P.~Eckle, T.T.~Truong; J.Phys.A {\bf 22} (1989) 4027
%
\end{references}
\end{document}